\documentclass[twocolumn]{aastex631}

\newcommand{\mkms}{$\rm ~km~s^{-1}$}
\newcommand{\kms}{$\rm km~s^{-1}$}

\def \NMgII {N_\text{Mg\textsc{ii}}}
\def \nMgII {n_\text{Mg\textsc{ii}}}
\def \NclMgII {N_{\mathrm{cl,Mg\textsc{ii}},i}}

\submitjournal{ApJ}

\shorttitle{Modeling Small-Scale CGM Structure}
\shortauthors{Hummels \& Rubin et al.}
\graphicspath{{./}{figures/}}
\usepackage{mathtools}

\begin{document}

\title{\textsc{CloudFlex}: A Flexible Parametric Model for the Small-Scale Structure of the Circumgalactic Medium}

\author[0000-0002-3817-8133]{Cameron B. Hummels}
\affiliation{TAPIR, California Institute of Technology, Pasadena, CA 91125, USA}
\author[0000-0001-6248-1864]{Kate H. R. Rubin}
\affiliation{Department of Astronomy, San Diego State University, San Diego, CA 92182 USA}
\affiliation{Center for Astrophysics and Space Sciences, University of California, San Diego, La Jolla, CA 92093, USA}
\collaboration{2}{These authors contributed equally to this work.}
\author[0000-0001-9735-7484]{Evan E. Schneider}
\affil{Department of Physics \& Astronomy and Pitt PACC, University of Pittsburgh, Pittsburgh, PA 15260, USA}
\author[0000-0003-3806-8548]{Drummond B. Fielding}
\affiliation{Center for Computational Astrophysics, Flatiron Institute, 162 5th Ave, New York, NY 10010, USA}
\affiliation{Department of Astronomy, Cornell University, Ithaca, NY 14853, USA}



\begin{abstract}

We present \textsc{CloudFlex}, a new open-source tool for predicting the absorption-line signatures of cool gas in galaxy halos with complex small-scale structure.  Motivated by analyses of cool material in hydrodynamical simulations of turbulent, multiphase media, we model individual cool gas structures as assemblies of cloudlets with a power-law distribution of cloudlet mass $\propto m_\mathrm{cl}^{-\alpha}$ and relative velocities drawn from a turbulent velocity field.  The user may specify $\alpha$, the lower limit of the cloudlet mass distribution ($m_\mathrm{cl,min}$), and several other parameters that set the total mass, size, and velocity distribution of the complex.  We then calculate the \ion{Mg}{2} $\lambda2796$ absorption profiles induced by the cloudlets along pencil-beam lines of sight.  We demonstrate that at fixed metallicity, the covering fraction of sightlines with equivalent widths $W_{2796}<0.3$ \AA\ increases significantly with decreasing $m_\mathrm{cl,min}$, cool cloudlet number density ($n_\mathrm{cl}$), and cloudlet complex size.  We then present a first application, using this framework to predict the projected $W_{2796}$ distribution around ${\sim}L^*$ galaxies.  We show that the observed incidences of $W_{2796}>0.3$ \AA\ sightlines within $10~\mathrm{kpc}<R_{\perp}<50~\mathrm{kpc}$ are consistent with our model over much of parameter space.  However, they are  underpredicted by models with $m_\mathrm{cl,min}\ge100M_{\odot}$ and $n_\mathrm{cl}\ge0.03~\rm cm^{-3}$, in keeping with a picture in which the inner cool circumgalactic medium (CGM) is dominated by numerous low-mass cloudlets ($m_\mathrm{cl}\lesssim100M_{\odot}$) with a volume filling factor ${\lesssim}1\%$.  When used to simultaneously model absorption-line datasets built from multi-sightline and/or spatially-extended background probes, \textsc{CloudFlex} will enable detailed constraints on the size and velocity distributions of structures comprising the photoionized CGM.

\end{abstract}



\section{Introduction} \label{sec:intro}


The circumgalactic medium (CGM) is the reservoir of gas surrounding a galaxy and extending to its virial radius and beyond.  Cool ($<10^4$ K), photoionized material is well-known to pervade circumgalactic environments \citep[e.g.,][]{Stocke2013,Peeples2014,Werk2014,Prochaska2017}.  Spectroscopy of background quasi-stellar object (QSO) sightlines passing close to foreground luminous galaxies reveals a cool phase traced by low-ionization metal transitions (e.g., \ion{Mg}{2}, \ion{Si}{2}) with a covering fraction close to unity \citep[e.g.,][]{BergeronStasinska1986,Churchill2000,Chen2010,Werk2013,Nielsen2013,Churchill2013,Kacprzak2013,
Dutta2020,Huang2021}.  Significant observational evidence implies that this material occupies very little volume ($\lesssim1\%$; \citealt{Stocke2013,Werk2014,Cantalupo2014,
Hennawi2015,Stern2016,FaermanWerk2023}), and galaxy formation theory requires that it coexists with a pervasive hot (${\sim}10^6$ K) phase around massive systems 
\citep[e.g., having halo masses $M_h \gtrsim 10^{11-12} M_{\odot}$;][]{Maller04,DekelBirnboim2006,Keres2009,Tumlinson2011,Nelson2013,Werk2016}.  However, the physics permitting this coexistence is not well understood, as the cool phase should be subject to disruption via hydrodynamic instabilities as it moves through and interacts with the hot medium \citep[e.g.,][]{Klein1994}.

This tension has inspired a series of theoretical efforts, beginning with the cloud-crushing simulations of, e.g., \citet[][]{Jones1994,HeitschPutman2009,Joung2012}, and \citet{SchneiderRobertson2017}.  These works imply that cool structures moving through a hot medium are destroyed on a timescale that increases with cloud size and scales inversely with velocity.  More recently, \citet{mccourt18} suggested that thermally-unstable hot circumgalactic material may instead fragment as it cools, remaining in pressure equilibrium, and resulting in a fine mist of cool cloudlets with sizes $<0.1$ pc that is comoving with the hot background.  Most apropos to the physical conditions of circumgalactic gas, \citet{gronke22} used three-dimensional hydrodynamical simulations of a turbulent, multiphase medium to study cool gas mass evolution and fragmentation, characterizing the resulting distribution of cloudlet masses.  This work identified a critical size threshold above which cool structures survive and grow for a given cloud temperature, density, and Mach number \citep[see also, e.g.,][]{Fielding:2020, Sparre20, Li20, Kanjilal21, Tan21, Abruzzo22}.

Alongside these highly-resolved theoretical studies, cosmological simulations have sought to produce a realistic CGM through a dynamic balance of galactic accretion and internal feedback processes.  Yet these simulations traditionally have underpredicted the overall covering fraction and column densities for cool gas structures relative to those observed \citep{Stinson2012, Hummels13}.  Recently, some studies have introduced enhanced spatial and mass resolution in the halos in these simulations, generating smaller, more numerous cool cloud structures and driving an increase in column density and covering fraction, bringing the results more in line with observational constraints \citep{Vandevoort19, Hummels19, Peeples19}.  Despite resolving scales as small as 500 pc, however, the cool gas properties in these computationally expensive simulations do not appear to be converged, although \citet{Ramesh23} find that some properties (e.g., CGM cold gas fraction) converge on slightly smaller scales.

While these theoretical studies represent significant progress in offering plausible explanations for the physical origins of the multiphase CGM, their predictions for the sizes, masses, and velocity distributions of cool structures have not yet been tested against observations.  Indeed, observational constraints on these quantities are impossible to obtain from spectroscopy of individual background QSOs due to the pencil-beam nature of this probe.  However, with the numbers of newly-discovered gravitationally-lensed QSO systems now growing rapidly \citep{Monier1998,Rauch2002,Ellison2004,Chen2014,zahedy16,rubin18a,Lemon2019,Zahedy2020,Dawes2022,Chan2023,He2023}, as well as the growing abundance of projected galaxy pairs permitting spatially-resolved absorption tomography \citep{Lopez2018,Rubin2018c,Peroux2018,Lopez2020,Tejos2021,Fernandez-Figueroa2022,Bordoloi2022,Afruni2023}, we are poised for a dramatic improvement in our ability to empirically constrain the morphology of the cool CGM.
 
In light of these growing observational constraints, we propose a new analytic model for describing the cool gas structures in the CGM, motivated by high resolution simulations of cloud crushing \citep[e.g.,][]{Gronke18,gronke20, Sparre20, Li20, Kanjilal21, Abruzzo22} and turbulent boxes \citep[e.g.,][]{gronke22, Das23}. Our simple model describes a single, spherical cool gas complex, a few kiloparsecs in size, composed of a multitude of cloudlets.  The model is characterized by several parameters to designate the cloudlets' masses, sizes, positions, metallicities, and turbulent velocity structure.  
The flexibility of the model enables us to directly and inexpensively explore how changes to the physical configuration of the absorbing gas manifest observationally through forward modeling.  This in turn will significantly enhance our ability to constrain cool CGM morphological properties from absorption-line datasets.

\begin{figure*}[ht]
\includegraphics[width=\textwidth,trim={3cm 2.5cm 0cm 2cm},clip]{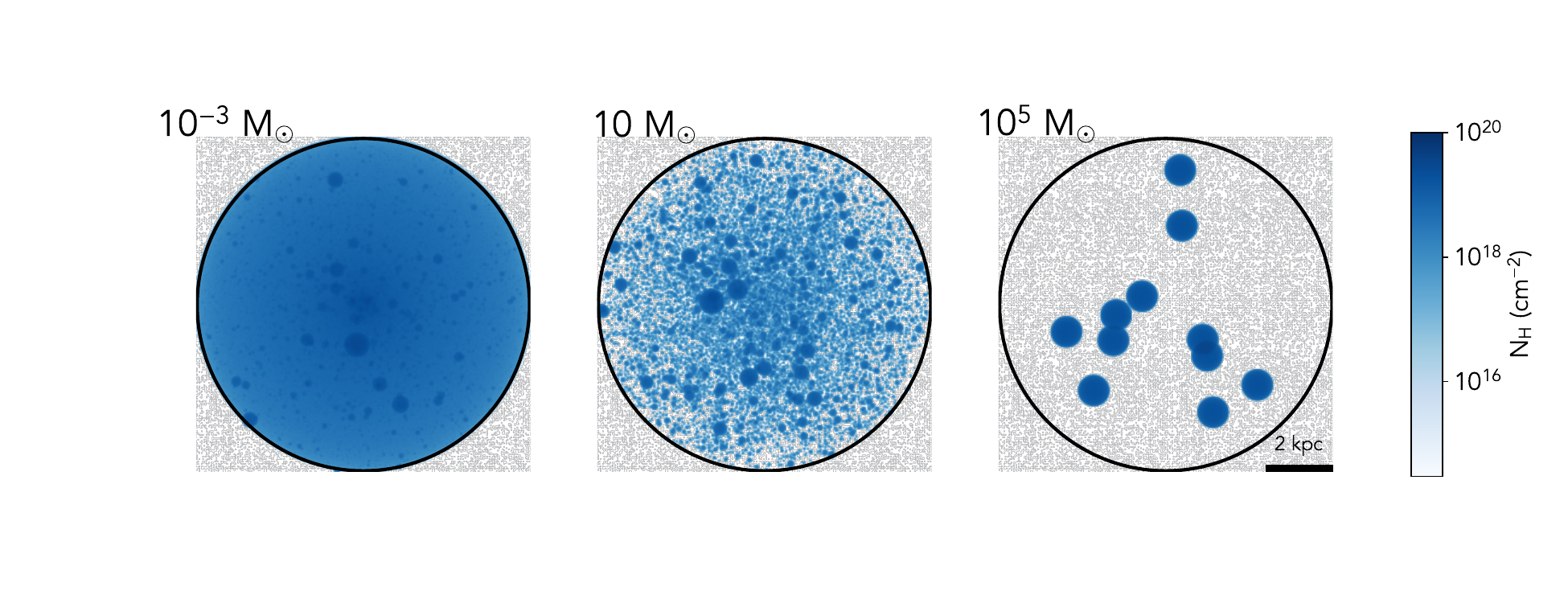}
\caption{Hydrogen column density projections for three realizations of our \textsc{CloudFlex} model, each of which represents an  assembly of cool cloudlets within a spherical volume with radius $d_{\rm cl, max}$.
The left-hand, middle, and right-hand panels show realizations with the minimum cloudlet mass set to $m_{\rm cl,min}=10^{-3}$, $10$, and $10^5 M_{\odot}$, respectively. 
All other parameters are set to their fiducial values (with $d_{\rm cl, max} = 5$ kpc, total complex mass $m_{\rm tot} = 10^6 M_{\odot}$, cloudlet mass function slope $\alpha = 2$, and cloudlet radial distance distribution slope $\zeta = 1$).  Each of the models shown has a volume filling factor $\approx 0.8$\%. 
\label{fig:demo}}
\end{figure*}

In Section \ref{sec:method}, we introduce the \textsc{CloudFlex} model\footnote{\url{https://github.com/cloudflex-project/cloudflex}}, its parameters, and the parameter values we adopt.  We describe a technique for producing synthetic observations from the model by sending sightlines through the cool gas distribution and generating absorption-line spectra using methods and source code from the \textsc{Trident} software package \citep{hummels17}.  
We use these synthetic spectra to predict
the distributions of \ion{Mg}{2} column densities and \ion{Mg}{2} $\lambda2796$ equivalent widths ($W_{2796}$) 
that would be observed toward background QSO sightlines for a given model.
In Section \ref{sec:analysis}, 
we vary 
the parameters of our model 
to investigate the impact of each on our synthetic observations, providing insight into which parameters have degenerate effects. 
In Section \ref{sec:discussion}, we 
describe a hypothetical CGM composed of many such cool gas complexes, predict its \ion{Mg}{2} absorption-line properties, and discuss the implications of these predictions in the context of observed circumgalactic $W_{2796}$ distributions.
In Section \ref{sec:conclusion},  we summarize our study and 
describe directions for future work using the \textsc{CloudFlex} model.

\section{Method} \label{sec:method}

\subsection{A Simple Model for a Circumgalactic Cool Gas Structure}
\label{subsec:model}

In this section we outline a simple parametric model for a cool gas structure (what we will call a ``complex'') in the halo of a galaxy.  A single complex is composed of many cool gas ``cloudlets" of varying masses and sizes occupying a spherical volume, and generated using a Monte Carlo method.  Cloudlet temperatures are assumed to be $10^4~\rm K$  and cloudlet velocities reflect turbulence in the surrounding hot ($10^6$~K) medium. All of the cool cloudlets are assumed to have a fixed gas density, which implies a uniform pressure over the region covered by a single cloudlet complex.  However, our model is agnostic as to the source of the pressure, as it may have contributions from thermal, magnetic, or cosmic ray sources. In this work we focus our attention on the observational signatures of the cool gas. Our ``fiducial'' cloudlet complex model is 5 kpc in radius with a total cool gas mass of $10^6M_{\odot}$.  
A galaxy's halo is nominally composed of many such structures, which together comprise its cool CGM.

Each complex is composed of cloudlets whose masses are drawn from a power-law distribution with slope $\alpha$:

\begin{equation} \label{eqn:mass}
    \mathrm{PDF}(m_{\rm cl}) = C_{\rm m} m_{\rm cl}^{-\alpha},
\end{equation}

\noindent with minimum and maximum cloudlet masses $m_{\rm cl, min}$ and $m_{\rm cl, max}$. We draw at random from this distribution until we have created a population with total mass $m_{\rm tot}$, which defines the constant of proportionality for the distribution $C_{\rm m}$. Similarly, we adopt a power-law distribution for the radial distance ($d_{\rm cl}$) of each cloudlet from the center of the complex: 

\begin{equation} \label{eqn:distance}
    \mathrm{PDF}(d_{\rm cl}) = C_{\rm d} d_{\rm cl}^{\:\zeta},
\end{equation}

\noindent with minimum and maximum distances $d_{\rm cl,min}$ and $d_{\rm cl,max}$, a slope $\zeta$, and $C_{\rm d}$ a constant of proportionality.  To place each cloudlet, we first draw a random distance $d_{\rm cl}$ from this distribution. The cloudlet's 3D location is then determined by 
selecting a random point with uniform probability on a sphere of radius $d_{\rm cl}$.  To do so, we draw random values for variates $u$ and $v$ from the range [0,1], with the azimuthal angle $\phi = 2\pi u$, and the polar angle $\theta = \cos^{-1}(2v - 1)$.  This ensures a uniform random distribution of cloudlet coordinates per area in solid angle.
The diameter of each complex is always equal to $2d_{\rm cl,max}$. 
In most of our models, we ensure that cloudlets do not overlap in physical space by testing the location of each new cloudlet against all others, and selecting a new location at random if needed.  This additional step can cause the final radial distribution of the cloudlets to differ from that of the specified PDF$(d_{\rm cl})$ for some values of $\zeta$, due to volume filling constraints.  
We describe this further in Section~\ref{subsubsec:clobber}.

Each cloudlet is assumed to have a uniform density $\rho_{\rm cl}$.  Thus, each cloudlet 
has a radius

\begin{equation} \label{eqn:radius}
    r_{\rm cl} = \left(\frac{3 m_{\rm cl}}{4\pi\rho_{\rm cl}}\right)^{1/3}.
\end{equation}
We convert between mass density, $\rho_{\rm cl}$, and number density, $n_{\rm cl}$, assuming a mean molecular weight $\mu=1$ (roughly appropriate for partially ionized gas):

\begin{equation}
    n_{\rm cl} = \frac{\rho_{\rm cl} }{{ m_{\rm p} \mu}},
\end{equation}
where $m_{\rm p}$ is the proton mass.   All cloudlets in a given complex are assumed to have the same metallicity $Z_{\rm cl}$.

Finally, we assume that turbulence in the surrounding hot medium imparts velocities to the cool cloudlets.  We extract individual cloudlet velocities from a turbulent velocity field generated from a power spectrum of the form

\begin{equation} \label{eqn:velocity}
    P_v(k) = C_{v} k^{-2\beta -1},
\end{equation}

\noindent where $k = d_{\rm cl}/\ell$ is the dimensionless wave number. Velocities drawn from this spectrum have a structure function (i.e., a root-mean-squared velocity difference as a function of 3D distance $\ell$) of the form
\begin{equation}
\label{eqn:turb}
    \langle (v(r) - v(r + \ell))^2 \rangle^{1/2} \propto \ell^{\beta},
\end{equation}
where a Kolmogorov spectrum has a value $\beta = 1/3$. We assume the turbulent driving is split between solenoidal and compressive modes at a 2:1 ratio. We sample the turbulent velocities over a grid of $N = [128, 128, 128]$ across the cloudlet complex, consequently limiting our wave numbers to ${\rm k_{min} = 1}$ and ${\rm k_{max} = 64}$.  We normalize the resulting cloudlet velocities with the parameter $v_{\rm max}$.

\begin{center}
\begin{table*}[tb!]
\begin{tabular}{ ||c | c | c | c|| }
\hline\
Parameter & Fiducial Value & Tested Range & Description\\ 
\hline\hline
$\alpha$ & 2 & $1.1-4$ & Cloudlet Mass Power Law Slope\\  
$m_{\rm cl, min}$ & $10 M_{\odot}$ & 10$^{-3}-10^{5} M_{\odot}$ & Minimum Cloudlet Mass\\
$m_{\rm cl, max}$ & $10^5 M_{\odot}$ & \nodata & Maximum Cloudlet Mass\\
$m_{\rm tot}$ & $10^{6}M_{\odot}$  & $10^{5}-10^{7} M_{\odot}$  &Total Mass in Cloudlets\\
$\zeta$ & $1$ & $-2- +5$ & Cloudlet Radial Distance Power Law Slope\\  
$d_{\rm cl, min}$ & 0.1 kpc & \nodata & Minimum Cloudlet Radial Distance\\
$d_{\rm cl, max}$ & 5 kpc & $2-11$ kpc & Maximum Cloudlet Radial Distance\\
$n_{\rm cl}$ & ${\rm 10^{-2}~cm^{-3}}$ & ${\rm 10^{-3} - 10^{-1}~cm^{-3}}$ & Cool Gas Number Density\\
$Z_{\rm cl}$ & $0.33 Z_{\odot}$  & $0.03-3.0 Z_{\odot}$ & Cool Gas Metallicity\\
$\beta$ & 0.33 & $0-1$ & Turbulent Power Spectrum Slope\\
$v_{\rm max}$ & 30 \kms & $8-100$ \kms & Turbulent Velocity Normalization\\
\hline
\end{tabular}
\caption{\textsc{CloudFlex} model parameters, their fiducial values, and explored ranges.}
\label{tab:params}
\end{table*}
\end{center}

\subsection{Model Parameter Values}

In Table~\ref{tab:params}, we present a complete list of our model parameters, 
the parameter values we choose for our ``fiducial'' complex model, and the range of each parameter explored in the following analysis. We emphasize that the goal of this work is not to develop quantitative constraints on these parameters via comparison to observations.
Rather, we study how systematically varying these parameters impacts observable quantities. While we adopt physically-motivated ranges for the values of our model parameters, they should not be interpreted as constraints arising from a fit to existing observations. 
Instead, we adopt these ranges to build a qualitative understanding of how observable quantities may be used to 
improve our constraints on the underlying physical distribution of cool circumgalactic gas.

\subsubsection{Cloudlet and Complex Masses: $\alpha$, $m_{\rm cl, min}$, $m_{\rm cl, max}$, $m_{\rm tot}$}

Each cool cloudlet has a mass, $m_{\rm cl}$, described by the power-law distribution defined in Equation~\ref{eqn:mass}.  For the slope of the power law, $\alpha$, we adopt a fiducial value of 2, motivated by the cloud mass distribution that has been found to arise in simulations of multiphase turbulent media \citep{gronke22,Das23}. The minimum cloudlet mass, ${m_{\rm cl,min}}$, is a primary variable in our analysis, as many recent studies have suggested that the CGM may contain vast numbers of small clouds that are currently unresolved in cosmological zoom simulations \citep{mccourt18,Hummels19,Vandevoort19,Suresh19,Ramesh23}. We adopt a fiducial minimum cloudlet mass of $10~M_{\odot}$, and test a range between $10^{-3}-10^{5}M_{\odot}$.  Here, the minimum bound is set in part to correspond to the weakest observed \ion{Mg}{2} column densities \citep[e.g.,][]{Churchill20}, and in part by computational memory constraints.  We use a maximum cloudlet mass of $10^5 M_{\odot}$ for all models in this study. 

We adopt a fiducial value for the total mass, $m_{\rm tot}$, of our cloudlet complex of $10^6 M_{\odot}$, and explore a range between $10^5 - 10^7 M_{\odot}$.
The analysis of \citet{Rubin2018c} of the coherence of \ion{Mg}{2} absorption equivalent width in the CGM of $L^*$ galaxies at $z\sim0.5$ implies a lower limit on their sizes of ${\gtrsim} 2$ kpc, which in turn implies masses ${\gtrsim} 10^4 M_{\odot}$ (assuming a solar abundance ratio, no dust depletion, and that \ion{Mg}{2} is the dominant ion).  On the other hand, high-velocity clouds (HVCs) in the halo of the Milky Way are known to have typical masses of ${\sim} 10^8 M_{\odot}$, including their ionized component \citep{LehnerHowk2011}.
The range of total complex masses we explore is therefore well within these observed constraints. 
Among the suite of models we generate, the total number of cloudlets per complex ranges from $10$ (for a model with $m_{\rm tot} = 10^6 M_{\odot}$ and  $m_{\rm cl,min} = 10^5 M_{\odot}$) to ${\sim} 5.5\times 10^7$ (for a model with $m_{\rm tot} = 10^6 M_{\odot}$ and  $m_{\rm cl,min} = 10^{-3} M_{\odot}$).

\subsubsection{Ensuring Cloudlets Do Not Overlap}
\label{subsubsec:clobber}
Because our model is fully analytic rather than an output from a hydrodynamical simulation, it is possible for some unphysical situations to arise.  In particular, because the locations of cloudlets are generated in a Monte Carlo fashion, cloudlets are permitted {\it ab initio} to occupy the same location in space and may overlap.
To prevent this scenario from arising, we implement an algorithm that 
avoids any cloudlet overlaps by testing the selected location of each new cloudlet to determine if it overlaps any existing cloudlets.  If an overlap is identified, a new random location for the cloudlet is assigned and retested.  This process is repeated until the cloudlet can be randomly placed in a vacant region, ensuring a physical distribution.
While preferable, this approach is very computationally expensive for models with ${>}10^6$ cloudlets due to its algorithmic inefficiency ($>O(N^2)$).  Thus, while we include this overlap test for most of the models in this study, we do not perform it for models with $m_{\rm cl,min} = 10^{-2} M_{\odot}$ and $10^{-3} M_{\odot}$, which include $6.5\times 10^6$ and $5.5\times 10^7$ cloudlets, respectively.

To qualitatively assess the potential systematic effect this limitation may have on our results, we use the approach outlined in Section~\ref{subsec:spectra} to calculate the \ion{Mg}{2} $\lambda 2796$ equivalent widths ($W_{2796}$) and column densities observed toward 10,000 background QSO sight lines for two versions of our $m_{\rm cl,min} = 10^{-1} M_{\odot}$ model: one constructed without cloudlet overlaps, and the other allowing overlaps.  Both models include
$8 \times 10^5$ cloudlets.  We find that the median and mean $W_{2796}$ values are equivalent to within $0.001$ \AA, though the maximum $W_{2796}$ is higher in the model that permits cloudlet overlap by $0.02$ \AA.  Both models have the same median and maximum numbers of cloudlets intercepted per sightline, and they have the same maximum \ion{Mg}{2} column densities to within 0.001 dex.  We conclude from this exercise that, for models in which all parameters are set at their fiducial values with the exception of $m_{\rm cl,min}$, permitting cloudlet overlap likely has only very minor effects on the resulting $W_{2796}$ and \ion{Mg}{2} column density distributions.  However, models in which overlaps are permitted allow for a higher density of cloudlets toward the center of the complex, which yields enhanced $W_{2796}$ values for sightlines passing close to the center.  This effect may become more pronounced as the number of cloudlets increases (i.e., for lower values of $m_{\rm cl,min}$).  We caution that the results we present below for the two models with $m_{\rm cl,min} = 10^{-2} M_{\odot}$ and $10^{-3} M_{\odot}$ should be interpreted with this caveat in mind.



\subsubsection{Cloudlet Radial Distribution: $\zeta$, $d_{\rm cl, min}$, $d_{\rm cl, max}$}\label{subsubsec:cloudlet_separation}

We have chosen a range in radii for the cloudlet complexes spanning $2~\mathrm{kpc} < d_{\rm cl, max} <11~\mathrm{kpc}$, with a fiducial value $d_{\rm cl,max} = 5$ kpc.  This is consistent with the lower limit on the coherence of \ion{Mg}{2} absorbers estimated by \citet[][${\gtrsim} 2$ kpc]{Rubin2018c}, and extends to slightly larger scales than the 95\% confidence constraint of $4.2^{+3.6}_{-2.8}$ kpc recently reported by \citet{Afruni2023}.
This interval also roughly spans the sizes of the more massive HVCs observed in the halo of the Milky Way, which can extend over several kiloparsecs \citep[e.g.,][]{Thom2008,DOnghiaFox2016}.  The so-called ``compact'' HVCs have median physical sizes of ${\sim} 10$ pc \citep{Putman2002,Saul2012}, and thus would be considered an individual cloudlet in our model.

We explore values of $\zeta$, the cloudlet radial distance power law slope, that span the range $-2 \le \zeta \le 5$.  However, we find that the inclusion of our test for cloudlet overlap results in radial distributions that differ from that specified for some values of $\zeta$, as it limits the numbers of cloudlets that are placed very close to the center of the complex.  This effect becomes stronger as the value of $\zeta$ decreases (i.e., as the distributions become more centrally concentrated).  
We find that for $\zeta=-2$, the final radial distance distribution is not described well by a single power law; instead, it is  better described by a broken power law that turns over at  $d_{\rm cl} \approx 1$ kpc.  
We choose $\zeta = +1$ as our fiducial value, such that the cloudlet volume density distribution declines approximately as $d_{\rm cl}^{-1}$, with the caveat that the final cloudlet distributions are likely to be modestly steeper in all models.  The extent to which the final distribution differs from $\zeta = +1$ may also have some dependence on the size of the complex ($d_{\rm cl, max}$) or the cloudlet sizes themselves (as set by their densities).  However, variations in the value of $\zeta$ over the range $0 \le \zeta \le 5$ have a negligible effect on the distributions of our primary observables ($W_{2796}$ and \ion{Mg}{2} column density), as shown in Appendix~\ref{app:other_params}.

\subsubsection{Cloudlet Thermodynamic Properties: $T_{\rm cl}$, $n_{\rm cl}$, $Z_{\rm cl}$}

An underlying assumption in our model is that all cool cloudlets have temperatures of $10^4$ K. At the cool gas densities we consider, the well-known combined effects of the collisional ionization equilibrium cooling function and photoionization by the ultraviolet (UV) background radiation should keep the gas close to this temperature. Photoionization modeling of UV absorption line data probing the CGM of low-redshift, ${\sim} L^*$ galaxies indicates that this temperature is indeed consistent with the absorption strengths of low-ionization species \citep[e.g.,][]{Stocke2013,Werk2014}, which are the focus of our observational comparisons (see Section~\ref{sec:discussion}). Cloudlet densities, on the other hand, are not as well constrained. 
This same photoionization modeling implies densities in the range $10^{-4.5}~\mathrm{cm^{-3}} \lesssim n_{\rm H} \lesssim 10^{-2}~\mathrm{cm^{-3}}$ in the low-redshift CGM  \citep[e.g.,][]{Stocke2013,Werk2014,Prochaska2017}, while weak \ion{Mg}{2} absorbers at intermediate redshifts have $10^{-3.5}~\mathrm{cm^{-3}} \lesssim n_{\rm H} \lesssim 10^{-0.5}~\rm cm^{-3}$ \citep{Rigby2002}.
In this work, we test a range of densities between $n_{\rm cl} = 10^{-3}~\rm cm^{-3}$ and $10^{-1}~\rm cm^{-3}$ and adopt a fiducial value of $n_{\rm cl} = 10^{-2}~\rm cm^{-3}$. We set the fiducial metallicity of cloudlets to $Z_{\rm cl} = \frac{1}{3} Z_\odot$, and explore a range between 0.03 and  3.0 $Z_\odot$, consistent with the range of constraints derived in \citet{Prochaska2017}, as well as with those found in the CGM of low-redshift ${\sim}L^*$ galaxies in the FIRE simulations \citep{Ji20}.

\subsubsection{Cloudlet Velocities: $\beta$, $v_{\rm max}$}

The velocity of each cloudlet is set by its 3D position within a grid of velocities generated by sampling a turbulence distribution. We choose a fiducial value of $\beta = \frac{1}{3}$ for the slope of the turbulent power spectrum, appropriate for subsonic turbulence (i.e., a Kolmogorov spectrum). We vary $\beta$ over a range from 0 to 1.0, where 0 corresponds to no correlation between distance and velocity, and 1.0 is an extremely steep correlation. While this range is larger than is theoretically expected, we adopt it in this work to demonstrate the relatively small impact of a changing slope on our observables (see Appendix~\ref{app:other_params}). We set the velocity normalization $v_{\rm max} = 30$ \kms\  for our fiducial model, and explore a range between 8 and 100 \kms, where the minimum represents an extremely low velocity dispersion across a structure that is several kiloparsecs in size, and the maximum is a value close to the velocity dispersion observed across entire halos \citep[e.g.,][]{Neeleman2013,Lau2016,UrbanoStawinski23}.\\

Figure \ref{fig:demo} illustrates the projected cloudlet distribution for three model complex realizations.  All parameters are set at their fiducial values with the exception of $m_{\rm cl, min}$, which is set to $10^{-3} M_{\odot}$, $10 M_{\odot}$, and $10^5 M_{\odot}$.  Each of these complexes contains the same total cool-phase mass ($m_{\rm tot} = 10^6 M_{\odot}$) and has the same volume filling factor ($v_{\rm ff} \approx 0.8\%$).  As expected, the figure demonstrates an increase in the projected surface area of cloudlets with decreasing $m_{\rm cl, min}$.

\begin{figure}[ht]
\includegraphics[width=\columnwidth,trim={4cm 6cm 4cm 6cm},clip=]{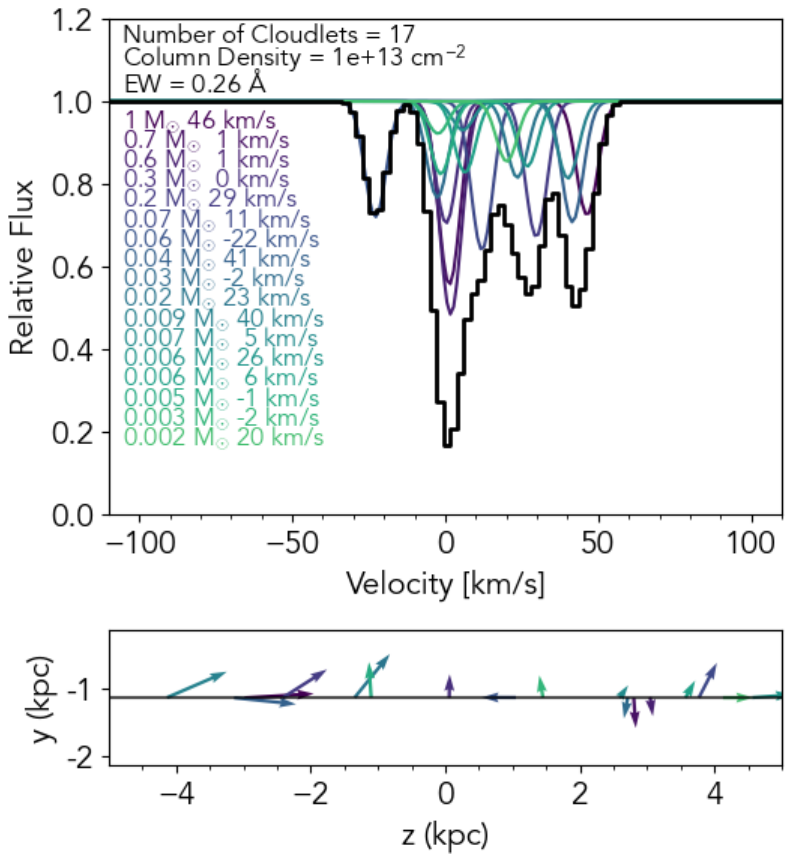}
\caption{{\it Top:} Simulated \ion{Mg}{2} $\lambda 2796$ absorption profile for a single sightline piercing a cloudlet complex with $m_{\rm cl, min}=10^{-3} M_{\odot}$. Colored curves represent the Voigt profile contribution of individual cloudlets, whose masses and line-of-sight velocities are printed in the same colors. The black histogram shows the resulting observed spectrum, smoothed to a spectral resolution $\mathcal{R} = 45000$. {\it Bottom:} The cross section of the sight line passing through the model complex volume with the observer on the left.  Each arrow represents an intersected cloudlet and its velocity vector in $y$-$z$ space.
\label{fig:spec_demo}}
\end{figure}

\subsection{Generation of Synthetic Absorption-Line Profiles}
\label{subsec:spectra}

A primary goal of our study is to investigate how absorption-line observables depend on the parameters of our model. 
Here, we focus on the observable properties of the \ion{Mg}{2} $\lambda 2796.35$ transition,  
due to the significant observational literature offering constraints on the distribution of \ion{Mg}{2} around galaxies \citep[e.g,][]{Chen2010,Werk2013,Nielsen2013,Kacprzak2013,Dutta2020,Huang2021}, along random QSO sight lines \citep[e.g.,][]{Churchill20}, and toward gravitationally-lensed sources \citep[e.g.,][]{Chen2014,Lopez2018,rubin18a,Lopez2020,augustin21,Afruni2023}.
However, the procedure we describe can in principle be adapted to predict absorption from any relevant ionic transition, by incorporating methods from the \textsc{trident} software package \citep{hummels17}.

The algorithm for producing mock spectra is as follows. 
We first generate a random location for an infinitesimal background source within the projected area of the model complex in the $x-y$ plane.   
We then identify which cloudlets are intersected by this sight line as it passes through the complex along the $z$-axis. For each intersected cloudlet ($i$), we calculate the pathlength of intersection $d\ell_i$, 
such that the total column density per cloudlet is
$N_{\mathrm{cl}, i} = n_{\mathrm{cl}} d\ell_i$.
We then use the cloudlet metallicity, $Z_{\rm cl}$, 
in combination with the solar abundance ratio of Mg, ($n_{\rm Mg}/n_{\rm H})_{\odot}$, and a \ion{Mg}{2} ionization fraction, $\nMgII/n_{\rm Mg}$, to compute a column density of \ion{Mg}{2} ions for the sightline passing through cloudlet $i$:

\begin{equation}
    \NclMgII = N_{\mathrm{cl}, i} \mu \mathrm{X} \left(\frac{Z_{\rm cl}}{Z_{\odot}} \right) \left( \frac{n_{\rm Mg}}{n_{\rm H}}\right)_{\odot} \left(\frac{\nMgII}{n_{\rm Mg}}\right), 
\end{equation}

\noindent where $\mathrm{X}=0.74$ is the hydrogen mass fraction \citep{Asplund2009}.
We adopt a value for the solar abundance of Mg $(n_{\rm Mg}/n_{\rm H})_{\odot} = 3.47\times10^{-5}$, as assumed in the \textsc{cloudy} photoionization modeling package \citep{Ferland1998,Holweger2001}. 
For the \ion{Mg}{2} ionization fraction, we adopt values derived from a series of single-zone \textsc{cloudy} ionization models assuming a Haardt-Madau UV background at $z=0.25$
available within the \textsc{trident} package \citep{hummels17}.
Over our range of $n_{\rm cl}$,  $\nMgII/n_{\rm Mg}$ is predicted to vary from 0.05 to 0.75.  For our fiducial model with $n_{\rm cl} = 0.01~\rm cm^{-2}$, we find $\nMgII/n_{\rm Mg} \approx 0.33$.  For reference, Table \ref{tab:properties} lists the sizes of individual cloudlets and their maximum hydrogen and \ion{Mg}{2} column densities as a function of mass, assuming our fiducial parameter values.

\begin{deluxetable}{lccc}
\tablecaption{Cloudlet Properties\label{tab:properties}}
\tablehead{
\colhead{$m_{\rm cl}$} & \colhead{$r_{\rm cl}$\tablenotemark{a}} & \colhead{$N_{\rm H, max}$\tablenotemark{a}\tablenotemark{b}} & \colhead{$N_{\rm Mg II, max}$\tablenotemark{a}\tablenotemark{b}}\\ 
$M_{\odot}$ & pc & cm$^{-2}$ & cm$^{-2}$ 
}
\startdata
$10^{5}$ & 500 & $2 \cdot 10^{19}$ & $8 \cdot 10^{13}$\\
$10^{4}$ & 200 & $1 \cdot 10^{19}$ & $4 \cdot 10^{13}$\\
$10^{3}$ & 100 & $5 \cdot 10^{18}$ & $2 \cdot 10^{13}$\\
$10^{2}$ & 50 & $2 \cdot 10^{18}$ & $8 \cdot 10^{12}$\\
$10^{1}$ & 20 & $1 \cdot 10^{18}$ & $4 \cdot 10^{12}$\\
$10^{0}$ & 10 & $5 \cdot 10^{17}$ & $2 \cdot 10^{12}$\\
$10^{-1}$ & 5 & $2 \cdot 10^{17}$ & $8 \cdot 10^{11}$\\
$10^{-2}$ & 2 & $1 \cdot 10^{17}$ & $4 \cdot 10^{11}$\\
$10^{-3}$ & 1 & $5 \cdot 10^{16}$ & $2 \cdot 10^{11}$\\
\enddata
\tablenotetext{a}{Calculated assuming fiducial model parameters (with $n_{\rm cl} = 0.01~\rm cm^{-3}$ and $Z_{\rm cl} = 0.33 Z_{\odot}$).} 
\tablenotetext{b}{$N_{\rm H, max}$ and $N_{\rm MgII, max}$ represent the total hydrogen and \ion{Mg}{2} column densities observed along sightlines 
passing through the cloudlet's center.}
\end{deluxetable}

We then use functions from the \textsc{trident} software package to deposit an absorption feature as a Voigt profile for each intercepted cloudlet, using its \ion{Mg}{2} column density $\NclMgII$, the velocity of the cloudlet along the $z$-axis ($v_{\mathrm{z},i}$), the degree of thermal broadening due to the gas temperature ($T_{\rm cl} = 10^4$ K), and the expected degree of turbulent broadening (described in Section~\ref{subsec:therm_and_turb}) as inputs.  We repeat this process for every cloudlet intersected by the sight line to calculate the superposition of all Voigt profiles.   
For this study, we apply a Gaussian convolution to smooth these ideal profiles to a spectral resolution of $\mathcal{R}=45000$ and sample them with a spectral dispersion of $2.2$\mkms, consistent with the fidelity of the Keck/HIRES \citep{Vogt1994} and VLT/UVES \citep{Dekker2000} high-quality QSO spectral databases used in the \ion{Mg}{2} analysis of \citet{Churchill20}.


Figure \ref{fig:spec_demo} shows an example of the resulting spectrum of a single sight line, with the lower panel representing the sidelong path of the sight line through the modeled cloudlet distribution with the observer on the left. Overplotted on the sightline are the intersected cloudlets with arrows representing their individual velocity vectors. The upper panel displays the corresponding absorption features for each cloudlet, color-coded to match the cloudlets in the bottom panel and labeled by their respective masses and line-of-sight velocities. The black channel map is the aggregate absorption spectrum that would be seen by an observer.  It is worth noting that despite the four clear kinematic components present in this spectrum, the underlying gas lacks any obvious kinematic or spatial coherence.

For each sight line and spectrum, we calculate the number of cloudlets intersected, the total \ion{Mg}{2} column density of gas traversed ($\NMgII$), and the equivalent width of the resulting \ion{Mg}{2} absorber ($W_{2796}$). We generate a sample of $10^4$ sightlines placed at random behind each cloudlet complex model and calculate these quantities for each sightline.  The resulting distributions of the number of cloudlets per sightline, $\NMgII$, and $W_{2796}$ for the fiducial model are shown in Figure~\ref{fig:histograms} (outlined in black). We also compute the covering fraction of the cloudlet complex ($f_{\rm C}^{\rm cc}({>}W_{\rm X})$), i.e., its fractional incidence of absorbers having strengths $W_{2796} > W_{\rm X}$ as a function of $W_{\rm X}$ among our $10^4$ sightline sample, and show these values in the right-most panels of Figure~\ref{fig:histograms}. The fiducial model exhibits $f_{\rm C}^{\rm cc}\approx0.4$ for equivalent width limits in the range $0.01-0.05$ \AA, and values of only $f_{\rm C}^{\rm cc}\approx0.15$ for $W_{2796} > 0.1$ \AA.



\subsubsection{Thermal and  Turbulent Broadening}\label{subsec:therm_and_turb}

The width of a spectral absorption feature depends on a combination of physical mechanisms.  In our model,
the thermal motions of the ions in each cloudlet contribute a broadening  $b\approx 2~\rm km~s^{-1}$ for $10^4$ K gas. Doppler broadening is represented by the addition of absorption features from individual cloudlets, accounting for the motion of multiple cloudlets projected into the same line of sight. Finally, turbulent broadening can in principle result from the velocity dispersion of the gas within a single cloudlet. 

We estimate this intra-cloud contribution by extending the turbulent velocity cascade described in Equation~\ref{eqn:turb} to sub-cloud scales. 
In a scenario in which this turbulence is generated by an ambient hot medium in pressure equilibrium with the cool cloudlets, the intra-cloud component may be estimated by assuming the conservation of kinetic energy between the two media.  This implies that the intra-cloud turbulent velocity $v_{\rm cl, turb}$, is equal to the product $ \chi^{-1/2} v_{\rm turb, hot}$, where $\chi$ is the ratio of densities of the cool and hot phases, and $v_{\rm turb, hot}$ is the turbulent velocity of the hot material.  



Using Equation~\ref{eqn:turb}, we first calculate the implied 3D turbulent velocity differences as a function of distance between two spatial locations within a cool cloudlet.  We then rescale this relation by the factor $\chi^{-1/2}$ to calculate the 3D turbulent velocities expected for the cool phase.  We adopt a value $\chi=100$, appropriate for a hot phase with $T = 10^6$ K in pressure equilibrium with our cloudlets at $T = 10^4$ K.
We cap the maximum value at the sound speed of the cool medium, $c_{\rm s} \sim 10$\mkms. We further assume that these 3D turbulent velocities are isotropic on average, and divide by $\sqrt{3}$ to estimate the 1D turbulent velocity broadening relation. 

For each cloudlet pierced by a background sightline, we use the intersected pathlength $d\ell$ as our distance, and calculate the implied 1D turbulent broadening as described above.
We add this value in quadrature with the thermal broadening, and adopt this sum as the total velocity broadening for the absorption feature from the cloudlet. 
We have found that because the intra-cloud turbulent broadening we calculate is almost always $< 3~\rm km~s^{-1}$ for our fiducial values $\beta=1/3$ and $v_{\rm max}=30~\rm km~s^{-1}$, its impact on our results is minor.  Nevertheless, we include this source of broadening in all subsequent analysis.

\subsection{Accessing the \textsc{CloudFlex} Code}
\label{subsec:code}

 \textsc{CloudFlex} is an open-source,
object-oriented, pure Python code.  Members of the scientific community are actively encouraged to use, develop, and collaborate with \textsc{CloudFlex}, and we release it according to the Revised BSD License. \textsc{CloudFlex} can be found on GitHub at \url{https://github.com/cloudflex-project/cloudflex}.


\begin{figure*}[ht]
\includegraphics[width=\textwidth]{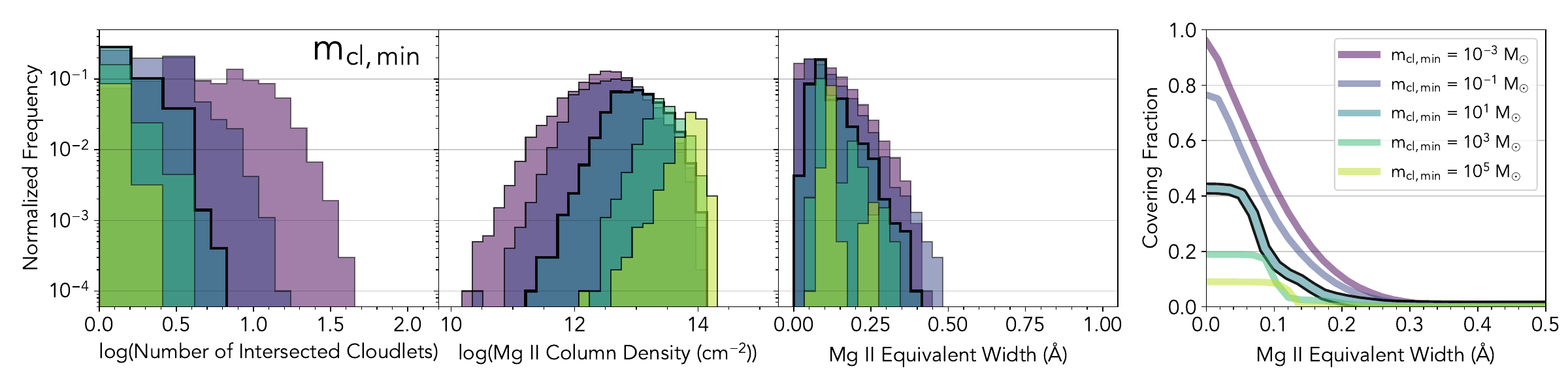}
\includegraphics[width=\textwidth]{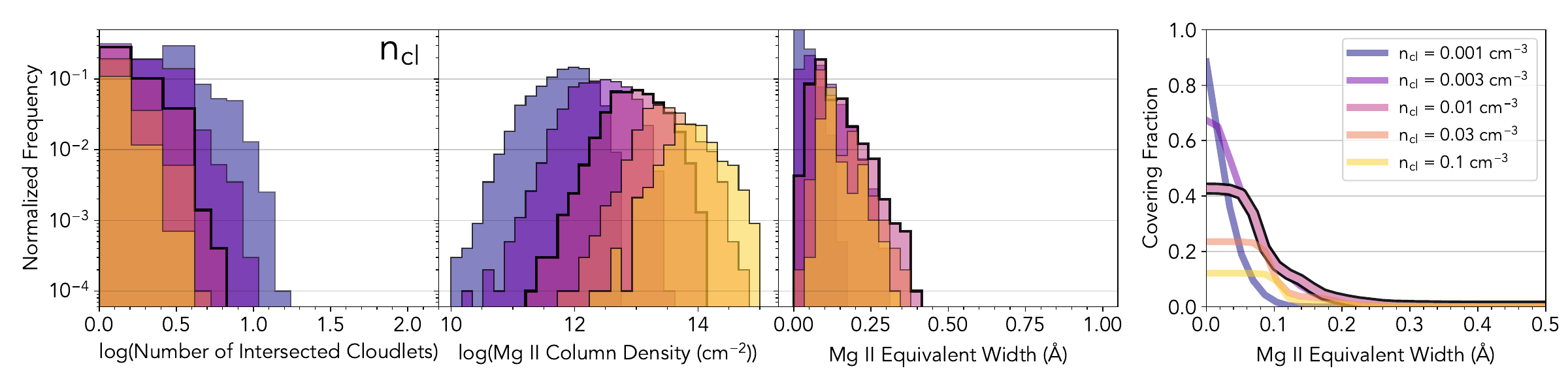}
\includegraphics[width=\textwidth]{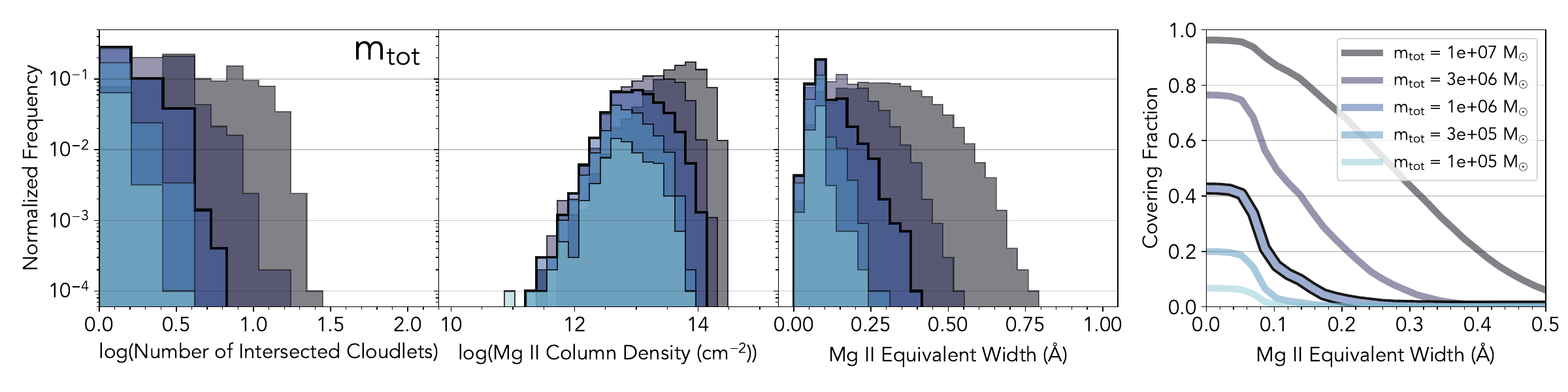}
\includegraphics[width=\textwidth]{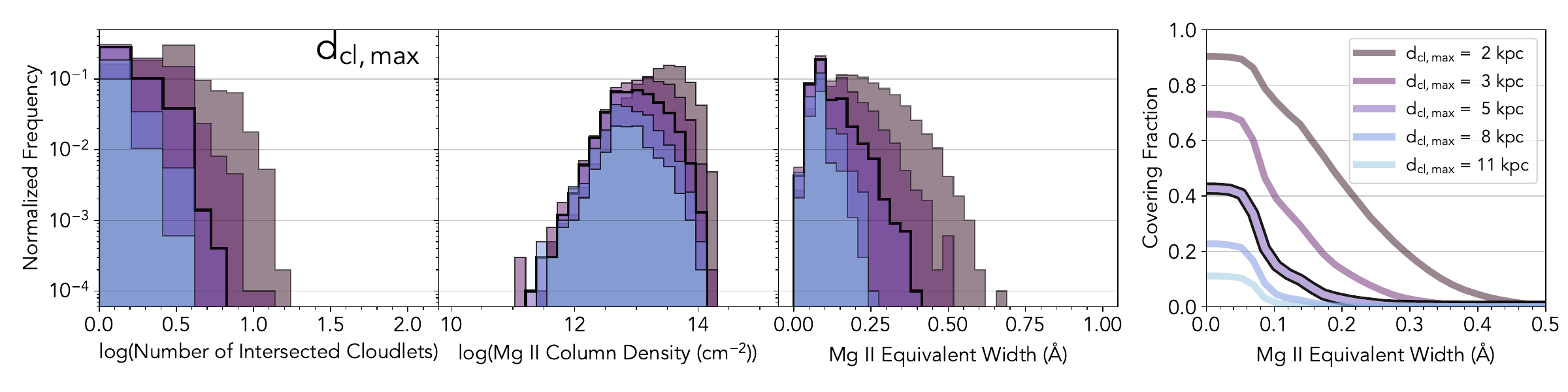}
\caption{Probability distributions of cloudlet complex \ion{Mg}{2}-absorption properties for models with varied input parameters. Each row shows results for variations in a single model parameter ($m_{\rm cl, min}$, $n_{\rm cl}$, $m_{\rm tot}$, and $d_{\rm cl,max}$, from top to bottom), with all other parameters fixed at their fiducial values. From left to right, each column shows distributions of the number of cloudlets intercepted per sightline, \ion{Mg}{2} column density ($\NMgII$), and \ion{Mg}{2} equivalent width $W_{2796}$, with the fiducial model outlined in black.  Each distribution is normalized by the number of sight lines sampled, and we have excluded $W_{2796}$ values of 0 \AA\ from all $W_{2796}$ distributions for clarity.  The right-most panels show the covering fraction of \ion{Mg}{2} absorption for equivalent widths greater than the values indicated on the $x$-axis ($f_{\rm C}^{\rm cc}({>}W_{\rm X})$).
\label{fig:histograms}}
\end{figure*}

\section{The Relation Between Cloudlet Complex Parameters and \ion{Mg}{2} Absorption}
\label{sec:analysis}

Here we present a preliminary investigation of the effects of varying several of our model parameters on the observables described above. As the goal of this work is to demonstrate the impact of changes in these parameters on observable quantities, rather than to fit our model to existing observations, we choose to vary only one parameter at a time, leaving all others at their fiducial values. This elucidates qualitative trends between model parameter values and observed \ion{Mg}{2} column densities, equivalent widths, and absorber covering fractions. We will present results from a more exhaustive exploration of parameter space in future work.

In the sections below, we describe the effects of varying four model parameters which we have found to have significant impacts on our observables: the minimum cloudlet mass $m_{\rm cl,min}$ (Section~\ref{subsec:analysis_mclmin}); the density of cool gas $n_{\rm cl}$ (Section~\ref{subsection:ncl}); the total mass of cloudlets $m_{\rm tot}$ (Section~\ref{subsec:total_mass}); and the radius of the cloudlet complex $d_{\rm cl, max}$ (Section~\ref{subsection:dclmax}).  
Changes in $\alpha$, $\zeta$, $Z_{\rm cl}$, $v_{\rm max}$, and $\beta$ have overall more modest effects on observed \ion{Mg}{2} column density and $W_{2796}$ distributions, which we demonstrate 
in Appendix~\ref{app:other_params}.

\subsection{Minimum Cloudlet Mass: $m_{\rm cl,min}$}\label{subsec:analysis_mclmin}

We present distributions of observables for cloudlet complexes having a range of $m_{\rm cl, min}$ between $10^{-3} M_{\odot}$ and $10^{5} M_{\odot}$ in the top row of panels in Figure~\ref{fig:histograms}.  Reducing the value of this parameter across this range results in a greater than order-of-magnitude increase in the maximum number of intersected cloudlets, as well as a significant broadening of the $\NMgII$ distribution to include much weaker absorbers.  The $W_{2796}$ distributions for those models with $10^{-3}M_{\odot} \le m_{\rm cl,min} \le 10 M_{\odot}$ appear similar in shape; however, the number of sightlines with $W_{2796}=0$ \AA\ declines significantly for smaller $m_{\rm cl,min}$, resulting in substantially higher covering fraction ($f_{\rm C}^{\rm cc}({>}W_{\rm X})$) values for $W_{\rm X} < 0.1$ \AA.  The $W_{2796}$ values for the model with $m_{\rm cl,min}=10^5 M_{\odot}$ have a multimodal distribution, with a strongly dominant peak at $0$ \AA, and a secondary peak at $0.10-0.15$ \AA\ resulting from sightlines that pass through a single cloudlet with mass $10^5 M_{\odot}$.  Note that we have excluded all sightlines with $W_{2796} = 0$ \AA\ from the $W_{2796}$ distributions for clarity, so the former peak is not shown.   The covering fractions of the two models with $m_{\rm cl,min} \ge 10^3 M_{\odot}$ are suppressed relative to those of the fiducial model at nearly all $W_{\rm X}$ limits, with the most significant offsets at $W_{\rm X} < 0.08$ \AA.

\begin{figure}[ht]
\includegraphics[width=\columnwidth]{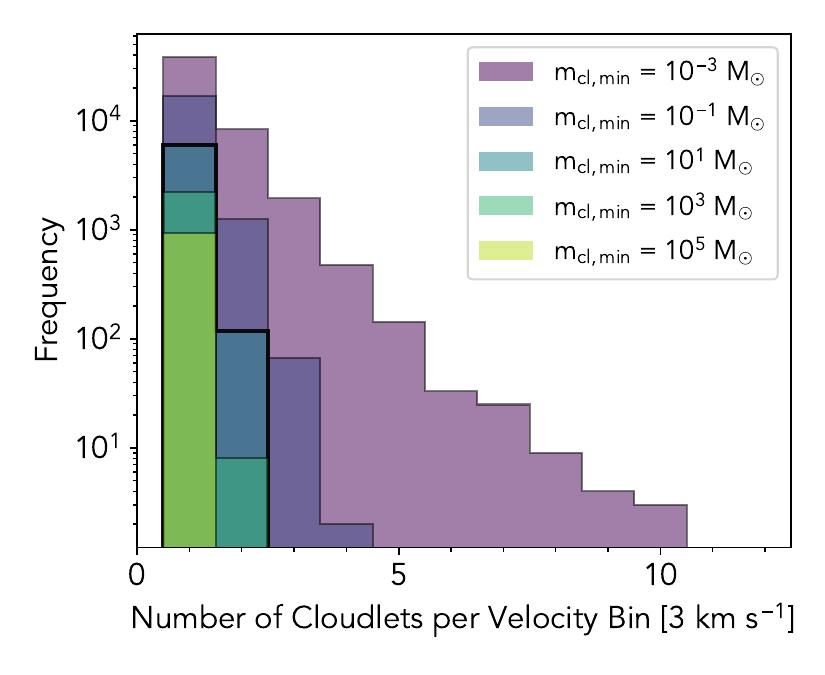}
\caption{Distribution of the number of cloudlets arising along the same background sight line within velocity bins of width $\Delta v = 3~\rm km~s^{-1}$ for complexes with varied $m_{\rm cl,min}$ (as indicated in the legend). Bins without cloudlets are excluded for clarity.
Cloudlets arising within $3~\rm km~s^{-1}$ of each other would be difficult to distinguish in traditional Voigt profile analyses, even at the high spectral resolution we simulate ($\mathcal{R}=45000$). 
\label{fig:nclouds_per_velbin}}
\end{figure}

To further explore the implications of the large numbers of intersected cloudlets arising along many of our simulated sightlines (in particular for models with $m_{\rm cl, min} \le 10 M_{\odot}$), we assess the degree to which these cloudlets overlap in velocity space.  We start by dividing the velocity space of intersected cloudlets for each simulated sight line into bins of width $\Delta v = 3~\rm km~s^{-1}$ (i.e., only slightly larger than our pixel width of $2.2~\rm km~s^{-1}$).  We then count the number of intersected cloudlets arising per bin.  Finally, we assemble these counts across all sight lines simulated for a given complex model, and show their distributions in Figure~\ref{fig:nclouds_per_velbin}.  

For all of these models, the vast majority of $3~\rm km~s^{-1}$ velocity bins contain only one cloudlet (if any).  However, as $m_{\rm cl,min}$ decreases to $0.001 M_{\odot}$, the relative frequency of velocity bins containing two or more cloudlets increases to ${\approx} 22\%$.  We emphasize that these ``overlapping'' cloudlets are physically distinct structures which may arise up to $10$ kpc from each other (see Figure~\ref{fig:spec_demo}).  Moreover, even at the spectral resolution of Keck/HIRES and VLT/UVES, it would be very difficult to isolate the absorption profiles intrinsic to such structures via traditional Voigt profile fitting \citep[e.g.,][]{Crighton2013,Crighton2015,Rudie19,Churchill20}.  \citet{Marra2022} drew similar conclusions from their analysis of absorption arising in cosmological hydrodynamic simulations of dwarf galaxy environments. Indeed, because most analyses adopt the minimum number of absorption  components required for an acceptable best spectral fit, they are likely insensitive to distinct structures arising at velocity separations significantly greater than $3~\rm km~s^{-1}$ \citep[e.g.,][]{Hafen23}. For example, the Voigt profile fits to the \ion{H}{1} and metal lines of Lyman Limit Systems described by \citet{Crighton2013,Crighton2015} and \citet{Zahedy21} report components that are separated by at least $10~\rm km~s^{-1}$ in every case.  We caution that if the cool material being probed in these systems has significant fine-scale structure as implied by the idealized hydrodynamical simulations upon which \textsc{CloudFlex} is based, a significant fraction of these components is likely to include absorption from multiple distinct cloudlets.

\subsection{Density of Cool Gas: $n_{\rm cl}$}\label{subsection:ncl}

Distributions of observables for cloudlet complex models with varied cool gas densities ($n_{\rm cl}$) are shown in the second row of panels in Figure~\ref{fig:histograms}.  
Lower $n_{\rm cl}$ values yield larger cloudlets with a higher volume filling factor and total projected area, leading to larger numbers of intersected cloudlets overall.  However, the adopted \ion{Mg}{2} ionization fraction also declines with $n_{\rm cl}$, such that the $\NMgII$ distributions shift toward overall lower values.  As the cloudlet density decreases from $n_{\rm cl} = 0.1~\rm cm^{-3}$ to $0.01~\rm cm^{-3}$, the maximum observed values of $W_{2796}$ increase to $\approx 0.40~\rm \AA$ due to the larger numbers of intersected cloudlets; however, as $n_{\rm cl}$ drops further, the incidence of absorbers with $W_{2796}\gtrsim0.06~\rm \AA$ begins to decrease due to the concomitant decreasing cloudlet \ion{Mg}{2} column densities.


\subsection{Total Mass of Cloudlet Complex: $m_{\rm tot}$}
\label{subsec:total_mass}

We show distributions of the number of intersected cloudlets, \ion{Mg}{2} column densities, and $W_{2796}$ for models with total complex masses varying from $m_{\rm tot} = 10^5 M_{\odot}$ to $10^7 M_{\odot}$ in the third row of panels in Figure~\ref{fig:histograms}.  Larger values of $m_{\rm tot}$ broaden all of these distributions, increasing the maximum number of intersected cloudlets by nearly an order of magnitude, and yielding increasing numbers of sightlines with  $\NMgII >10^{14}~\rm cm^{-2}$.  Larger values of $m_{\rm tot}$ likewise yield larger $W_{2796}$ values up to $\approx 0.8$ \AA, with significantly larger covering fractions at all $W_{\rm X}$.  

\subsection{Radius of Cloudlet Complex: $d_{\rm cl, max}$}\label{subsection:dclmax}

Finally, we show distributions of observables for cloudlet complexes with variable maximum cloudlet radial distances ($d_{\rm cl,max}$) in the bottom row of Figure~\ref{fig:histograms}.  For each model shown, the locations of the background sightlines are modified such that they are distributed across the projected area $\pi d_{\rm cl,max}^2$.  Larger values of $d_{\rm cl,max}$ therefore result in fewer intersected cloudlets per sightline, lower maximum column densities, and lower $W_{2796}$ values.  The smallest value explored, $d_{\rm cl,max}= 2$ kpc, yields a covering fraction profile higher across all $W_{\rm X}$ than the model with $m_{\rm tot}=3\times10^6 M_{\odot}$ and $d_{\rm cl,max}=5$ kpc shown in the panel above. Changes to both $m_{\rm tot}$ and $d_{\rm cl,max}$ imply significant changes in the complex
volume filling factor $v_{\rm ff}$. 
The similarity of the trends seen in the third and fourth rows of Figure \ref{fig:histograms} 
demonstrate that these parameters affect our \ion{Mg}{2} observables in a similar manner.

From this exploration, it is clear that the distributions of these observables for individual cloudlet complexes
do not depend on individual parameter values in ways that are uniquely identifiable.
We proceed with extensions of this analysis under the assumption that direct comparison to observed CGM properties may rule out specific combinations of the model parameters.  We pursue a first example of such a comparison in the following section.

\section{Initial Implications}
\label{sec:discussion}

\begin{figure*}[ht]
\includegraphics[width=\textwidth,trim={0.3cm 0.3cm 0cm 0cm},clip]
{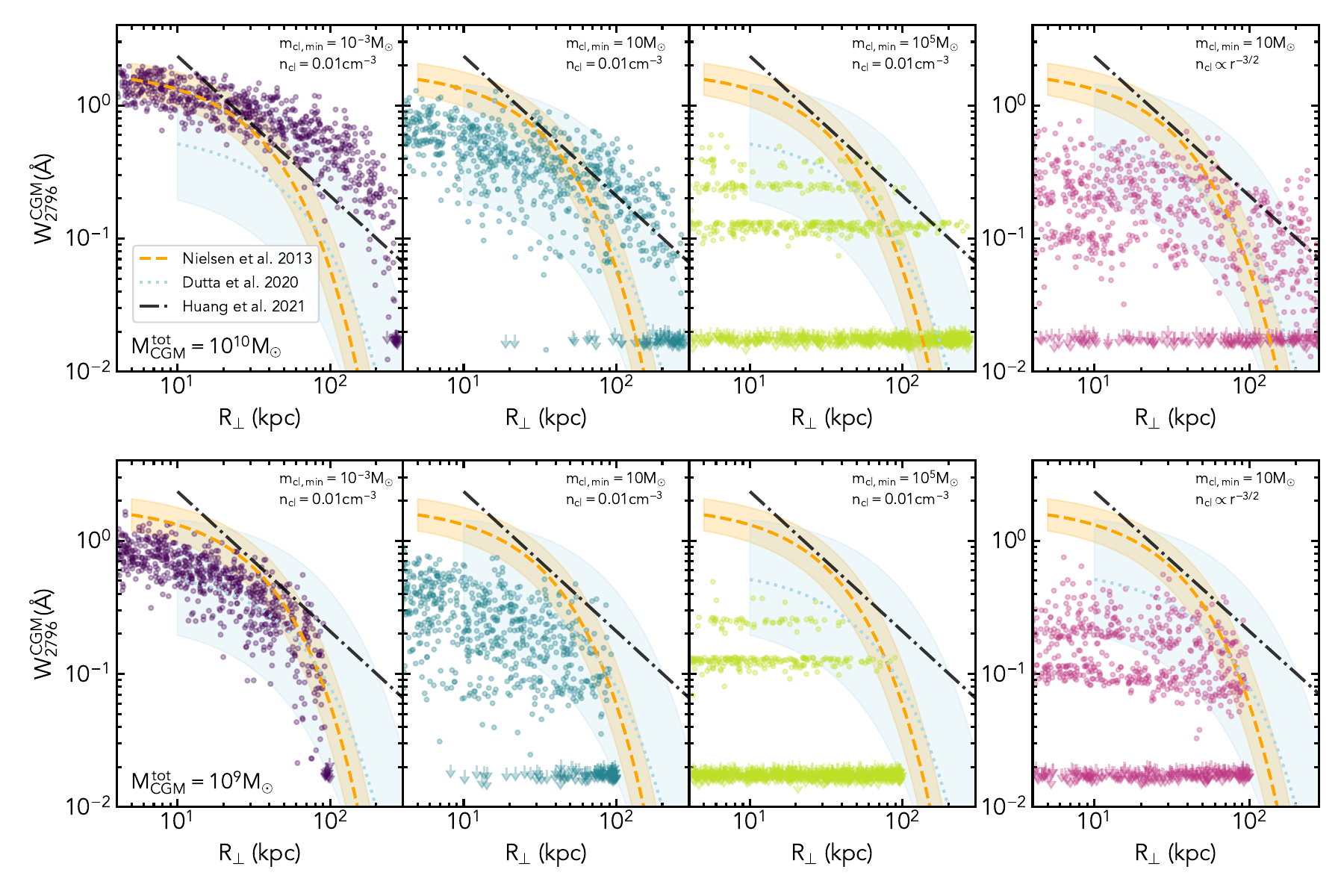}
\caption{Maximum \ion{Mg}{2} $\lambda 2796$ equivalent widths predicted along QSO lines of sight passing within halos containing a total cool CGM mass $M_{\rm CGM}^{\rm tot} = 10^{10} M_{\odot}$ (top row) and $M_{\rm CGM}^{\rm tot} = 10^{9} M_{\odot}$ (bottom row). For predictions shown in the first three columns, cool material is distributed in cloudlet complexes 
having minimum cloudlet masses $m_{\rm cl,min} = 10^{-3} M_{\odot}$, $10 M_{\odot}$, and $10^5 M_{\odot}$, respectively (with all other parameters set at their fiducial values).  
The right-most panels show the $W_{2796}^{\rm CGM}$ distributions predicted for a model in which the cloudlet density in the complexes varies as $n_{\rm cl} = 0.1~\mathrm{cm^{-3}} (20~\mathrm{kpc}/R)^{3/2}$.  
Best-fit models for the observed relation between $W_{2796}$ and $R_{\perp}$ reported by \citet{Nielsen2013}, \citet{Dutta2020}, and \citet{Huang2021} are shown in orange, light blue, and black, respectively. The \citet{Nielsen2013} and \citet{Huang2021} galaxy samples have median halo masses $\log M_h/M_{\odot}\sim 11.7-12$, while the \citet{Dutta2020} galaxy sample has a median halo mass $\log M_h/M_{\odot}\sim 11.5$.
The contours show $\pm1\sigma$ uncertainties in the \citet{Nielsen2013} and \citet{Dutta2020} relations.  
Under these assumptions, the $m_{\rm cl,min} = 10^{-3} M_{\odot}$ model is broadly consistent with the observed $W_{2796}$ profiles within $R_{\perp}<50$ kpc.
\label{fig:Rperp_EW}}
\end{figure*}

The $W_{2796}$ values predicted by our model are not directly comparable to those observed toward background QSO sightlines probing \ion{Mg}{2} absorption in the CGM of foreground galaxies.  In reality, the CGM must be composed of many cloudlet complexes, such that a given background sightline may probe numerous absorbing structures along its path through the halo.  

\subsection{Building a Halo-Scale Model}

To generate a rough estimate of the number of cloudlet complexes that a sightline at a given $R_\perp$ might intercept, we construct a toy CGM in which the mass in the cool photoionized phase is distributed as 

\begin{equation}
    M_{\rm CGM} (<R) = M_{\rm CGM}^{\rm tot} \left ( \frac{R}{R_{\rm vir}} \right )^2,
\end{equation}

\noindent with $M_{\rm CGM}^{\rm tot}$ the total cool gas mass within the virial radius, $R_{\rm vir}$.  This distribution is inspired by the findings of \citet{Stern2016}, who constructed a model that represents the photoionized CGM as collections of small, dense cloudlets embedded in larger, lower-density clouds.  They found that this multi-density model is consistent with the ionic column densities measured in the COS Halos survey \citep{Werk2013} for $M_{\rm CGM}^{\rm tot} = (1.3 \pm 0.4) \times 10^{10} M_{\odot}$, with cloud sizes that scale as $\sim \rho^{-1}$, and with cloud densities that harbor \ion{Mg}{2} having sizes of $\approx 50$ pc.  

Our model differs from that of \citet{Stern2016} in that we assume that all of the mass is in the $10^4$ K phase of the CGM ($M_{\rm CGM}^{\rm tot}$) and is distributed in cloudlet complexes with a given $m_{\rm tot}$, $\alpha$, $m_{\rm cl,min}$, $m_{\rm cl,max}$, etc.  By design, these cloudlets therefore have a range of physical sizes at a given density $n_{\rm cl}$.  We divide our toy CGM into a series of $n$ spherically-symmetric shells of width $\Delta R$  with outer radii $R_i$ and mass $\Delta M(R_i)$, and calculate the number of complexes needed to populate the $i$th shell, $N_{\mathrm{cc},i} = 
\Delta M(R_i)/m_{\rm tot}$, as well as $n_{\mathrm{cc},i}$, the number of cloudlet complexes per unit volume in each shell $i$.


The cross-sectional area of each cloudlet complex is simply $\sigma_{\rm cc} = \pi d_{\rm cl,max}^2$. The number of cloudlet complexes intercepted along a given sightline is then $N_{\rm cc, sightline} = \int n_{\rm cc}(R) \sigma_{\rm cc} ds$, with $ds$ representing the line element.  For our discrete CGM shells, we may compute this integral as

\begin{align*}
    N_{\rm cc, sightline}(R_\perp) = 2 \sigma_{\rm cc} n_{\mathrm{cc}, imin} \sqrt{R_{imin}^2 - R^2_\perp} \\ 
    + 2 \sigma_{\rm cc} \sum_{i=imin+1}^{imax} n_{\mathrm{cc},i} \sqrt{R_{i}^2 - R^2_\perp} - \sqrt{R_{i-1}^2 - R^2_{\perp}}, 
\end{align*}

\noindent with $imin$ indicating the innermost shell for which $R_{\perp} < R_i$, and $imax$ indicating the outermost shell.



In a physical halo, cloudlet complex-like structures would trace the bulk kinematic motions of the cool material, and their absorption features would likely have overlapping line-of-sight velocities.  In this case, the observed equivalent width would be less than the sum of the equivalent widths arising from all intersected complexes.  We do not account for this overlap here, and instead simply estimate an upper limit to the observed equivalent widths ($W_{2796}^{\rm CGM}$) by drawing $N_{\rm cc, sightline}(R_{\perp})$ values of $W_{2796, j}$ at random from the $W_{2796}$ distribution of the corresponding cloudlet complex model (as shown in Figure~\ref{fig:histograms}) and calculating $W_{2796}^{\rm CGM} = \sum_{j} W_{2796,j}$.  We emphasize that these predictions represent the maximum absorption strength that could potentially be observed along each sightline in this scenario.

We show the results of this exercise for our fiducial cloudlet complex model, as well as two variants of the fiducial model with $m_{\rm cl,min}=10^{-3}M_{\odot}$ and $m_{\rm cl,min} = 10^5 M_{\odot}$, in Figure~\ref{fig:Rperp_EW}.  We assume a total cool CGM mass of $M_{\rm CGM}^{\rm tot} = 10^{10} M_{\odot}$ and a virial radius $R_{\rm vir} = 280$ kpc (consistent with the best-fit $M_{\rm CGM}^{\rm tot}$ reported by \citealt{Stern2016} and their adopted $R_{\rm vir}$) for all predictions shown in the top panels, and 
adopt $M_{\rm CGM}^{\rm tot} = 10^{9} M_{\odot}$ and $R_{\rm vir} = 100$ kpc for predictions shown at bottom.  We
adopt a CGM shell width of $\Delta R = 10$ kpc in all cases.  We draw 1000 values of $R_{\perp}$ at random from a distribution that is uniform in $\log R_{\perp}$, and use these values to compute a corresponding distribution of $W_{2796}^{\rm CGM}$ for each cloudlet complex model (indicated in the upper right of each panel).  
$W_{2796}^{\rm CGM}$ values $<0.01$ \AA\ are indicated as upper limits at $\approx 0.02$ \AA. 
We also compute the implied volume filling factor $f_{\rm V, cc}$ of the complexes by assuming that the volume of each individual complex is $V_{\rm cc} = 4\pi d_{\rm cl, max}^3 / 3$.  We find that 
the central CGM shell (i.e., the region within $R<10$ kpc) is slightly over-filled for both values of $M_{\rm CGM}^{\rm tot}$, with $f_{\rm V, cc} = 1.6$ and 1.25 for $M_{\rm CGM}^{\rm tot} = 10^{10} M_{\odot}$ and $10^{9} M_{\odot}$, respectively.  We do not consider this to be strictly unphysical, as the cloudlet complexes themselves have a filling factor of only $v_{\rm ff} \approx 0.8\%$.  However, we caution that this scenario runs contrary to our assumption that all cool cloudlet complexes in a given CGM realization have the same geometry (i.e., that they are spatially distinct from each other).  The values of $f_{\rm V, cc}$ in the central-most shells are also somewhat sensitive to the choice of shell width $\Delta R$; however, the effect of varying $\Delta R$ is negligible for $R_{\perp}\gtrsim 20$ kpc, where $f_{\rm V, cc} \lesssim 0.3$ for both values of $M_{\rm CGM}^{\rm tot}$.

This simple exercise demonstrates explicitly that 
under the numerous assumptions laid out above, 
the covering fraction of strong \ion{Mg}{2} absorbers (with $W_{2796}>0.3~\rm \AA$) is strongly dependent on the minimum cloudlet mass adopted in our cloudlet complex model for a given $M_{\rm CGM}^{\rm tot}$.  Whereas the $M_{\rm CGM}^{\rm tot} = 10^{10} M_{\odot}$ model with $m_{\rm cl, min}=10^5 M_{\odot}$ yields a halo covering fraction ($f_{\rm C}^{\rm halo}$) of absorbers having $W_{\rm 2796}^{\rm CGM} > 0.3~\rm \AA$ of only $0.1$ within $R_{\perp} < 50$ kpc, the halo covering fractions increase to $f_{\rm C}^{\rm halo}({>}0.3~\mathrm{\AA}) = 0.86$ and $1.0$ for $m_{\rm cl, min}=10 M_{\odot}$ and $m_{\rm cl, min}=10^{-3} M_{\odot}$, respectively.  
For the $M_{\rm CGM}^{\rm tot} = 10^{9} M_{\odot}$ models, the  $f_{\rm C}^{\rm halo}({>}0.3~\mathrm{\AA})$ values within $R_{\perp} < 50$ kpc are 0.92, 0.48, and 0.03 for $m_{\rm cl, min} = 10^{-3}M_{\odot}$, $10M_{\odot}$, and $10^5M_{\odot}$.  We also find that the structure in the $W_{2796}$ distribution for the $m_{\rm cl,min}=10^5 M_{\odot}$ cloudlet complex model is reflected in the uneven distribution of $W_{2796}^{\rm CGM}$ values shown in Figure~\ref{fig:Rperp_EW}: 
 $53\%$ of the non-zero equivalent widths exhibited by this model are in the range $0.10-0.15$ \AA. Most of our line-of-sight equivalent widths are approximate multiples of these values in this case.


\subsection{A Halo-Scale Model with Varying Cool Cloudlet Density}

Each of our previous realizations assume a constant $n_{\rm cl} = 0.01~\rm cm^{-3}$ for all complexes, regardless of their location relative to the halo center.  However, the few available observational constraints on the density profiles of hot material in galaxy halos imply that they decline with radius \citep[e.g.,][]{AndersonBregman2011,Dai2012,MillerBregman2013,MillerBregman2015}.  Moreover, cooling flow solutions to the steady-state equations that model the hot CGM as an ideal fluid imply that the gas number density $n_{\rm H} \propto R^{-3/2}$ \citep{Stern2019}.  In a scenario in which the temperature of this material varies only weakly with $R$, and where our $T\sim 10^4$ K cloudlets are in thermal pressure equilibrium with the hot phase, 
the gas number density of the cloudlets would likewise decline with distance from the halo center as $n_{\rm cl} \propto R^{-3/2}$.  

To assess the effect of such a dependence on the predictions shown in the first three panels of Figure~\ref{fig:Rperp_EW}, we construct a new toy halo with $M_{\rm CGM}^{\rm tot} = 10^{10}M_{\odot}$ and $R_{\rm vir} = 280$ kpc, again adopting a CGM shell width of $\Delta R = 10$ kpc.  For each shell, we calculate a cool cloudlet number density according to $n_{\mathrm{cl},i} = 0.1~\mathrm{cm^{-3}} (20~\mathrm{kpc}/R_i)^{3/2}$.  We then assign a different cloudlet complex model to each shell, drawing from the subset of models shown in the second row of Figure~\ref{fig:histograms} (i.e., those with $n_{\rm cl} = 0.001, 0.003, 0.01, 0.03$, and $0.1~\rm cm^{-3}$), and selecting the model with the $n_{\rm cl}$ value closest to $n_{\mathrm{cl},i}$.  For each simulated QSO sightline, we calculate its path length  through each shell, use this quantity to compute the number of cloudlet complexes intercepted per shell, and then draw that number of $W_{2796}$ values at random from the appropriate $W_{2796}$ distribution.  The final $W_{2796}^{\rm CGM}$ in this case is the sum of the equivalent widths predicted to arise per shell.
We show the values predicted for 1000 random values of $R_{\perp}$ in the right-most panels of Figure~\ref{fig:Rperp_EW}, calculating this distribution for both $M_{\rm CGM}^{\rm tot}=10^{10} M_{\odot}$ (top) and $M_{\rm CGM}^{\rm tot}= 10^9 M_{\odot}$ (bottom).  We find that  $W_{2796}^{\rm CGM}$ is suppressed in the former model at $R_{\perp} < 100$ kpc relative to those predicted for the constant $n_{\rm cl}=0.01~\rm cm^{-3}$, $m_{\rm cl,min}=10 M_{\odot}$ case, and that the magnitude of this effect is most prominent at small $R_{\perp}$. Absorption observed toward the halo center is dominated by that arising from the $n_{\rm cl}=0.1~\rm cm^{-3}$ model, which yields 
few absorbers with low overall equivalent widths (Figure~\ref{fig:histograms}).  Absorption in the halo outskirts is dominated by the higher covering fractions of the $n_{\rm cl} = 0.001-0.003~\rm cm^{-3}$ models.  We conclude from this exercise that models accounting for a decline in CGM density as a function of $R$ will tend to have flatter $W_{2796}^{\rm CGM}$ profiles overall, and will also have an overall lower incidence of $W_{2796}^{\rm CGM} > 0.3~\rm \AA$ absorbers relative to those that assume our fiducial value of $n_{\rm cl} = 0.01~\rm cm^{-3}$ held constant across the CGM.



\subsection{Comparison of Halo-Scale Model with Observations}

For comparison to these predictions, we refer to the \citet{Nielsen2013}, \citet{Dutta2020}, and \citet{Huang2021} studies of circumgalactic \ion{Mg}{2} absorption observed toward background QSOs.  \citet{Nielsen2013} performed a literature search to assemble the MAGIICAT database, which includes 182 QSO-galaxy pairs within $5.4~\mathrm{kpc} \le R_{\perp} \le 194$ kpc with spectroscopic coverage of the \ion{Mg}{2} transition over the redshift range $0.07 \le z \le 1.1$.  All of these galaxies were determined to 
 be isolated (i.e., they have no neighbor within a projected distance of 100 kpc and within a velocity offset of $\delta v <500 ~\rm km~s^{-1}$).  The galaxy sample spans a broad range of $B$-band luminosities ($0.02 \le L_B/L_B^* \le 5.87$), but has a median halo mass of $\log M_h/M_\odot = 12.01$ \citep{Churchill2013} and a median redshift $\langle z \rangle = 0.359$.  Thus, while this sample is quite diverse, it has median properties similar to that of the COS Halos galaxy sample at slightly lower redshifts ($\langle z \rangle =0.359$ vs.\ $z\sim0.2$).
\citet{Nielsen2013} performed extensive analysis of this dataset, finding a best-fit log-linear relation 
\begin{equation*}
    \log W_{2796}~\mathrm{(\AA)} = (0.27 \pm 0.11) + (-0.015 \pm 0.002) R_\perp.
\end{equation*}
\noindent This relation is shown in Figure~\ref{fig:Rperp_EW} with a dashed orange curve and contours.  We truncate the relation at the minimum impact parameter included in the dataset.


More recently, \citet{Huang2021} reported results from a large survey of \ion{Mg}{2} absorption within impact parameters $9~\mathrm{kpc} \le R_{\perp} \le 497$ kpc of 211 isolated galaxies.  The galaxies span a redshift range $0.10 < z< 0.48$ and have a median redshift $\langle z \rangle =0.21$.  The sample size permitted analysis of the dependence of $W_{2796}$ as a function of numerous galaxy properties (e.g., quenched status, $B$-band magnitude, stellar mass).  Considering only the isolated star-forming galaxies, which span a stellar mass range $10^8 < M_*/M_{\odot} < 10^{11.3}$, \citet{Huang2021} reported a best-fit log-log model 
\begin{align*}
    \log W_{2796}~\mathrm{(\AA)} = (1.42 \pm 0.25) + (-1.05 \pm 0.17) \log R_\perp \\
    + (0.21 \pm 0.08) \times (M_* - \tilde{M_*}), 
\end{align*}

\noindent with $\log \tilde{M_*}/M_{\odot} = 10.3$. We show this relation for $M_* = \tilde{M_*}$ with a dot-dashed black curve in Figure~\ref{fig:Rperp_EW}. The median stellar mass of the \citet{Huang2021} isolated star-forming galaxy sample is $\log M_*/M_{\odot} = 10.2$, which implies a median halo mass $\log M_h/M_{\odot} \sim 11.7$ \citep{Moster2013}.

\citet{Dutta2020} analyzed \ion{Mg}{2} absorbers associated with galaxies discovered in the MUSE Analysis of Gas around Galaxies (MAGG) spectroscopic survey.  This star-forming galaxy sample has a range in redshift $0.8 < z < 1.5$, and spans stellar masses between $10^7 < M_*/M_\odot < 10^{12}$ with a median $\langle M_*/M_{\odot} \rangle = 2\times 10^9$. 
The ``typical'' MAGG galaxy has a  halo mass of $3\times 10^{11} M_{\odot}$ with $R_{\rm vir} = 100$ kpc at $z=1$.  For those absorbers that are associated with more than one galaxy, \citet{Dutta2020} used the match with the highest $M_*$ to avoid double-counting sightlines.  They found a best-fit relation
\begin{equation*}
    \log W_{2796}~\mathrm{(\AA)} = -0.20^{+0.42}_{-0.39} + (-0.009 \pm 0.003) R_{\perp}.
\end{equation*}
\noindent We likewise show this relation with a dotted light blue curve and contours in Figure~\ref{fig:Rperp_EW}.

We have tailored our choices of $M_{\rm CGM}^{\rm tot}$ for the predictions above to be appropriate for comparison to these datasets.  
We find that models with $M_{\rm CGM}^{\rm tot} = 10^{10} M_{\odot}$ exhibit upper limits on $W_{2796}^{\rm CGM}$ values well below those of the MAGIICAT and \citet{Huang2021} samples within $R_{\perp} < 40$ kpc for both $m_{\rm cl,min} = 10M_{\odot}$ and $10^5 M_{\odot}$.  Our $m_{\rm cl,min} = 10^{-3} M_{\odot}$ model is more consistent with the observations at $R_{\perp} = 10-40$ kpc, but exhibits maximum $W_{2796}^{\rm CGM}$ values which are ${\gtrsim}0.5$ \AA\ those observed at $R_{\perp} > 60$ kpc.  A comparison between the models constructed with $M_{\rm CGM}^{\rm tot} = 10^{9} M_{\odot}$ and the \citet{Dutta2020} relation yields similar results: those models with $m_{\rm cl,min} \ge 10 M_{\rm cl,min}$ exhibit $W_{2796}^{\rm CGM}$ values that lie below those observed (though many of the predicted values lie within the $1\sigma$ uncertainties in the observed relation at $m_{\rm cl,min} = 10 M_{\odot}$), whereas the model with $m_{\rm cl,min} = 10^{-3}M_{\odot}$ is broadly consistent with the observed relation.  

\begin{figure*}[ht]
\includegraphics[width=\textwidth,trim={0cm 0cm 0cm 0cm},clip]{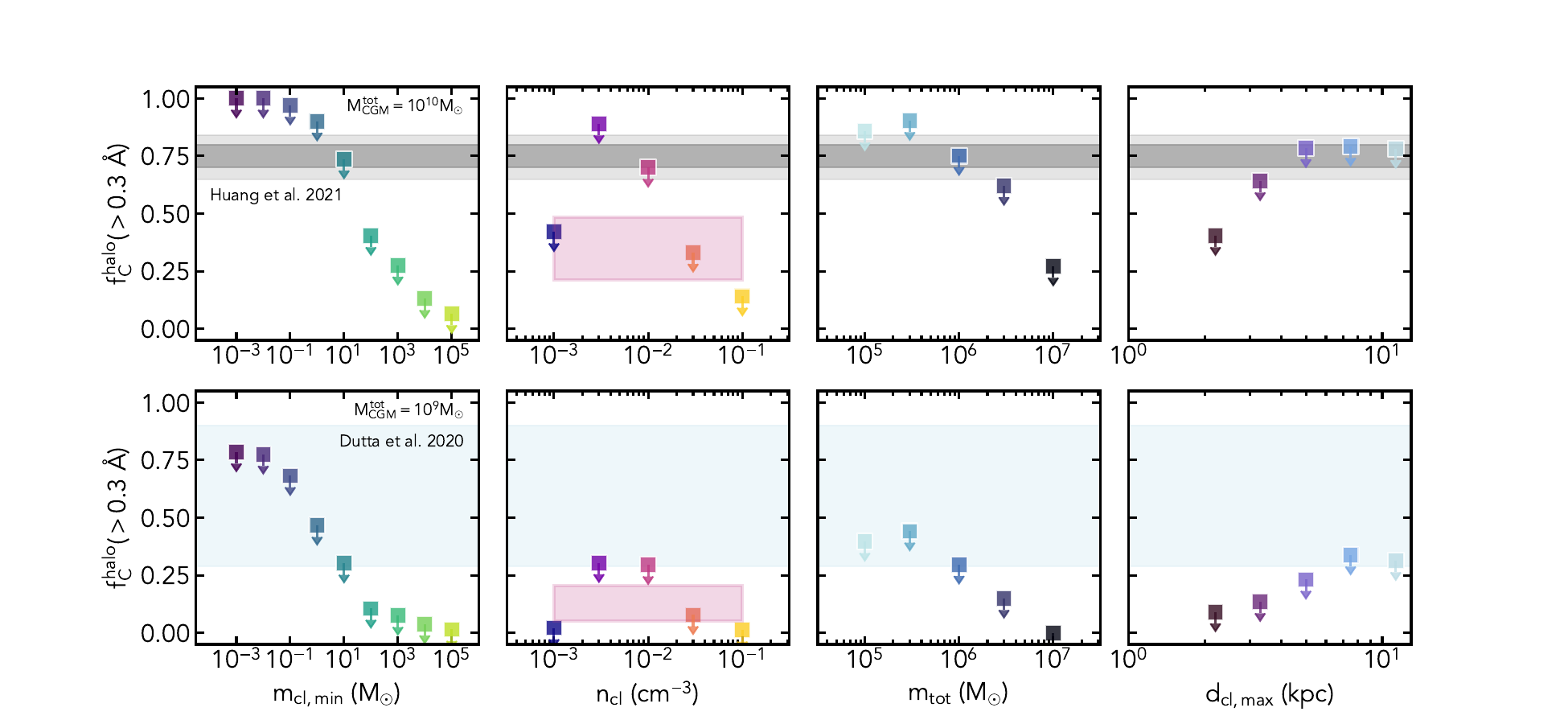}
\caption{Maximum predicted incidence of sight lines with $W_{\rm 2796}^{\rm CGM} > 0.3$ \AA\ within $10~\mathrm{kpc} < R_{\perp} < 50$ kpc in halos simulated as described in Section~\ref{sec:discussion} as a function of $m_{\rm cl,min}$, $n_{\rm cl}$, $m_{\rm tot}$, and $d_{\rm cl, max}$. Colors for each model correspond to those of the distributions shown in Figure~\ref{fig:histograms}. The top and bottom panels show predictions for halos with $M_{\rm CGM}^{\rm tot} = 10^{10} M_{\odot}$ and $10^{9} M_{\odot}$, respectively. The rose-colored bars show the range in $f_{\rm C}^{\rm halo}({>}0.3~\rm \AA)$ recovered in models in which the $n_{\rm cl}$ in the cloudlet complexes declines with $R$.
The dark and light gray bars show the $\pm1\sigma$ and $\pm2\sigma$ confidence intervals for the covering fraction of $>0.3$ \AA\ absorbers measured within $R_{\perp} < 50$ kpc of the \citet{Huang2021} isolated, star-forming galaxy sample. The   
light blue bars show the $\pm1\sigma$ confidence interval measured for the covering fraction of $>0.3$ \AA\ absorbers within $R_{\perp} < 50$ kpc by \citet{Dutta2020}.}
\label{fig:cf}
\end{figure*}

\subsection{Implications of Halo-Scale \ion{Mg}{2} Equivalent Widths}

At face value, these results imply that minimum cloudlet masses $\lesssim 10 M_{\odot}$ are required to reproduce the large equivalent width values (${\gtrsim} 0.3$ \AA) observed within $R_{\perp}<50$ kpc of luminous galaxies.  However, the inconsistency between the observations and the $m_{\rm cl, min} = 10^5 M_{\odot}$ model evident in Figure~\ref{fig:Rperp_EW} may also in principle be resolved in numerous other ways: e.g., by increasing $M_{\rm CGM}^{\rm tot}$, or by lowering the $n_{\rm cl}$ of the cloudlets to increase their internal covering fractions.  We leave a full investigation of these degeneracies to future work.

We have repeated the procedure described above to construct $W_{2796}^{\rm CGM}$ profiles for each of the cloudlet complex models generated as described in Section~\ref{sec:analysis}.  
We summarize these predictions in 
Figure~\ref{fig:cf}, which shows the incidence of sight lines with $W_{2796}^{\rm CGM} > 0.3$ \AA\  measured within $10~\mathrm{kpc} < R_{\perp} < 50$ kpc for each profile ($f_{\rm C}^{\rm halo}(\rm {>}0.3~\AA)$).  The dark and light gray bars indicate the $\pm1\sigma$ and $\pm2\sigma$ Wilson score confidence intervals for the  covering fraction of ${>}0.3$ \AA\ absorbers measured within $R_{\perp} < 50$ kpc of the isolated star-forming galaxies in the \citet{Huang2021} sample, and the light blue bars indicate $\pm1\sigma$ confidence intervals for
comparable covering fractions reported by \citet{Dutta2020}.  \citet{Nielsen2013} reported a $\pm1\sigma$ covering fraction confidence interval of $\sim0.80-0.88$ for $W_{2796} > 0.3$ \AA\ absorbers within 50 kpc of the full MAGIICAT sample, consistent with the \citet{Huang2021} constraint.
We also include results for the $n_{\rm cl} \propto R^{-3/2}$ toy halo, represented by the rose-colored bar.  We have found that these results are somewhat sensitive to the choice of $\Delta R$, and use the bar to show the range in $f_{\rm C}^{\rm halo}(\rm >0.3~\AA)$ values calculated for $\Delta R=10-20$ kpc.  We further caution here that covering fraction measurements are sensitive to the observational sampling across the relevant interval in $R_{\perp}$.   The datasets discussed here include relatively few sightlines at $R_{\perp} < 30$ kpc, whereas the sampling in our model is dominated by these small impact parameters.  Predicted covering fractions that account more precisely for the $R_{\perp}$ sampling in each of these datasets would therefore likely yield yet lower values than are shown in Figure~\ref{fig:cf}.

The models we have generated with $m_{\rm cl,min} \ge 10^2 M_{\odot}$, $n_{\rm cl} \ge 0.03~\rm cm^{-3}$ and $n_{\rm cl}=0.001~\rm cm^{-3}$, $m_{\rm tot} \ge 3\times 10^6 M_{\odot}$, and $d_{\rm cl,max} \le 2$ kpc yield $f_{\rm C}^{\rm halo}(\rm {>}0.3~\AA)$ values inconsistent with those observed.  Stated more precisely, we rule out models of the cool CGM in which it is composed entirely of cloudlet complexes with the geometry we have defined, and with, e.g., $m_{\rm cl,min} \ge 10^2 M_{\odot}$ and $n_{\rm cl}=0.01~\rm cm^{-3}$, $m_{\rm tot} = 10^6 M_{\odot}$, and $d_{\rm cl, max} = 5$ kpc.
This does not imply that we may rule out all scenarios in which, e.g., $m_{\rm cl, min} \ge 10^2 M_{\odot}$.  
On the contrary, our models adopt numerous simplifying assumptions regarding the physical conditions (e.g., a constant Mg abundance and \ion{Mg}{2} ionization fractions drawn from single-zone \textsc{cloudy} models), morphology, and total mass of the cool CGM.  Even within this framework, there is significant flexibility in this model that has not yet been explored.  For example, cloudlet complexes with $m_{\rm cl,min} = 100 M_{\odot}$ may be brought into accord with observations in several different ways, including by decreasing the adopted $n_{\rm cl}$, decreasing the adopted $m_{\rm tot}$, or by increasing the overall size of the complexes ($d_{\rm cl,max}$). 

On the other hand, our results imply that it would be difficult to reconcile models 
adopting $m_{\rm cl,min}\ge 10^2 M_{\odot}$ and having either high or very low cool phase densities ($n_{\rm cl} \ge 0.03~\rm cm^{-3}$ or $n_{\rm cl} \le 0.001~\rm cm^{-3}$), high total complex mass values ($m_{\rm tot} \ge 3\times 10^6 M_{\odot}$), and/or small complex sizes ($d_{\rm cl,max} \le 2$ kpc) with the observed \ion{Mg}{2}-absorbing CGM.  As shown in Figure~\ref{fig:appendix}, increasing the adopted Mg abundance or changing the values of the parameters governing the turbulent velocities of the cloudlets ($v_{\rm max}$, $\beta$) 
or their radial distance power law slope ($\zeta$) will likewise fail to increase the predicted $W_{2796}^{\rm CGM}$ values sufficiently to match those observed.  
Increasing the total cool CGM masses adopted by a factor of two improves agreement for models with $100 M_{\odot} \le m_{\rm cl,min} \le 1000 M_{\odot}$, $n_{\rm cl} \le 0.03~\rm cm^{-3}$, $m_{\rm tot} = 10^7 M_{\odot}$, and $d_{\rm cl,max} = 2$ kpc, but models with greater $m_{\rm cl,min}$ and $n_{\rm cl}$ remain in tension.  Increasing $M_{\rm CGM}^{\rm tot}$ by a factor of five brings the predicted covering fractions for all models shown into accord with those observed; however, it also requires volume filling factors $f_{\rm V, cc} \sim 1-7$ within $R < 40$ kpc for the fiducial model, and thus cannot be accommodated in our model framework.  

Referring to Table~\ref{tab:properties}, cloudlets with masses $\lesssim 100 M_{\odot}$ have
sizes $\lesssim 50$ pc for our fiducial parameter choices.  Recent theoretical
studies of the sizes of cloudlets capable of surviving
in a hot wind or turbulent flow have shown that the minimum size is set by the requirement that
hot gas from the wind must be able to mix with the cloud material and cool before
it advects past the cloud \citep[][see also \citealt{gronke22} for a slightly different physical interpretation but similar quantitative predictions]{Abruzzo2023, Tan23}.  This implies a criterion
$\alpha~r_{\rm cl} v_{\rm w} \gtrsim \tau_{\rm cool}$, where $v_{\rm w}$ is the relative
velocity of the cloudlet, $\tau_{\rm cool}$ is the characteristic cooling time, and
$\alpha$ is a scaling constant in the range $1 \lesssim \alpha \lesssim 10$ \citep{Abruzzo2023}.
For our adopted physical conditions, with $T_{\rm cl} = 10^4$ K, a density
ratio $\chi = 100$ between the cool and hot phases, a thermal pressure $=100~\rm K~cm^{-3}$,
and assuming the velocities of the cloudlets
are approximately equal to our fiducial turbulent velocity of $30~\mathrm{km~s^{-1}}$ (corresponding to a hot phase turbulent Mach number of $\mathcal{M}\approx0.2$),
Equation 11 of \citet{Abruzzo2023} implies a minimum survival radius of $\approx 50$ pc (comparable to the prediction using the \citealt{gronke22} model).
It has been shown in high resolution simulations of multiphase turbulence that above the minimum survival radius, the cloudlet mass distribution tends to follow a power-law with slope of 2 \citep{gronke22, Tan23, Fielding:2023}. The huge range of spatial scales inherent to this problem have, however, thus far made it infeasible to reliably simulate the  cloudlet mass distribution on scales well below minimum survival radius. It is, therefore, unclear from a theoretical standpoint how the mass distribution might change below this threshold. Our findings provide useful insight into this question since they imply that a significant population of cloudlets below the survival threshold are needed to reproduce the observations. 

Cloudlets with masses $\lesssim 100 M_{\odot}$ are likewise impossible to resolve
in state-of-the art cosmological zoom simulations \citep[e.g.,][]{Grand2017,Hopkins2018,Oppenheimer2018b}.  Even simulations that
pursue enhanced spatial resolution in the CGM \citep[e.g.,][]{Hummels19,Vandevoort19,Peeples19,Ramesh23}
are currently limited to resolutions of ${\gtrsim} 500$ pc or $1000 M_{\odot}$, and have not
yet demonstrated convergence of the mass distribution of cold CGM clouds \citep{Ramesh23}.
Given our suggestion that cloudlet masses may be distributed well below this limit (in combination
with the theoretical analyses that inspired our work; e.g., \citealt{mccourt18,Sparre20,gronke22}), such structures may never be fully resolved in a cosmological context due to computational limitations.

\section{Conclusions and Future Directions}
\label{sec:conclusion}

We have introduced \textsc{CloudFlex}, a new open-source tool to predict the absorbing properties of cool, photoionized CGM material with small-scale structure that may be flexibly defined by the user.  We employ a Monte Carlo method to model an individual cool gas structure as an assembly of cloudlets with a mass distribution specified by a power law and lower and upper limits $m_{\rm cl,min}$ and $m_{\rm cl,max}$.  The relative velocities of the cloudlets are drawn from a turbulent velocity field generated by a power spectrum with a variable power-law slope.  The cloudlet locations relative to the center of the complex are likewise drawn from a power law distribution with an outer limit $d_{\rm cl, max}$.  The user may specify several additional parameters, including the cloudlet gas number density, metallicity, and the total mass of the cloudlet complex.  A galaxy's cool CGM would be composed of many such cloudlet complexes.

We generate several realizations of this model, performing an introductory exploration of parameter space relevant to the cool CGM and guided by analyses of the mass evolution and fragmentation of cold clouds in idealized hydrodynamic simulations of turbulent, multiphase media.  We then calculate the \ion{Mg}{2} $\lambda 2796$ absorption line profiles induced by each model realization toward a large sample of randomly-placed background QSO sight lines. 
From this exercise we deduce the following:

\begin{itemize}
    \item At fixed metallicity, the covering fraction of \ion{Mg}{2} absorption of a given strength ($W_{\rm X}$) increases monotonically as the total complex mass ($m_{\rm tot}$) increases; as the overall complex size ($d_{\rm cl,max}$) decreases; and as the minimum cloudlet mass ($m_{\rm cl,min}$) decreases.  The dependence of the absorption covering fraction on the number density of the cloudlets ($n_{\rm cl}$) is more complex, but increases monotonically for $W_{2796}\lesssim 0.02$ \AA\ absorbers with decreasing $n_{\rm cl}$. 
    \item Variations in $v_{\rm max}$, the factor by which we normalize the velocity distribution of the cloudlets, 
    over the range $8~\mathrm{km~s^{-1}} \le v_{\rm max} \le 100~\rm km~s^{-1}$,
    can cause the maximum observed $W_{2796}$ to vary over the range $0.3-0.6$ \AA; however, variations in this parameter do not significantly affect the incidence of weaker absorbers ($W_{2796}<0.3$ \AA).
    Variations in the turbulent power spectrum slope over the range
    $0 \le \beta \le 1$ have a relatively minor effect on the observed $W_{2796}$ distributions.
    \item Many of our model realizations yield sightlines which intersect between 10 and 50 cloudlets (i.e., those with small minimum cloudlet masses in the range $10^{-3}~M_{\odot} \le m_{\rm cl,min} \le 10^{-1}~M_{\odot}$; low cool gas densities in the range  $0.001~\mathrm{cm^{-3}} \le n_{\rm cl} \le 0.003~\mathrm{cm^{-3}}$; large total masses in the range $3\times10^6 M_{\odot} \le m_{\rm tot} \le 10^7 M_{\odot}$; and small complex sizes in the range $2~\mathrm{kpc} \le d_{\rm cl,max} \le 3$ kpc).  Because cloudlets have relative velocities that fall in a limited range ($-60~\mathrm{km~s^{-1}} \lesssim \delta v \lesssim 60~\mathrm{km~s^{-1}}$), the resulting absorption troughs include contributions from numerous overlapping Voigt profiles arising from distinct structures.  Individual absorption components identified via traditional Voigt profile fitting of such systems will therefore include contributions from numerous cloudlets, rather than arising from unique, physically-separated clouds (see Figures~\ref{fig:spec_demo} and \ref{fig:nclouds_per_velbin}).  
\end{itemize}


We use this framework to predict the projected $W_{2796}$ distribution around ${\sim} L^*$ galaxies by making the assumption that their photoionized CGM is comprised of numerous such cloudlet complexes, and that its mass is distributed as $M_{\rm CGM}({<}R) \propto (R/R_{\rm vir})^2$.  Because we assume the complexes do not overlap in velocity space, our predictions represent the maximum $W_{2796}$ that can arise from each complex distribution.  
We compare these predictions to the $W_{2796}$ profiles observed at $z\sim0.2-0.3$ by \citet{Nielsen2013} and \citet{Huang2021} and at $z\sim1$ by \citet{Dutta2020}.
We show that the incidences of $W_{2796}>0.3$ \AA\ sightlines ($f_{\rm C}^{\rm halo}({>}0.3~\rm \AA))$ within $10~\mathrm{kpc} < R_{\perp}<50$ kpc are consistent with those observed over much of our parameter space.  However, they are underpredicted and inconsistent with the 95\% confidence interval for the covering fraction observed by \citet{Huang2021} in models with $m_{\rm cl, min} \ge 100 M_{\odot}$, 
$n_{\rm cl} \ge 0.03~\rm cm^{-3}$ or $n_{\rm cl} \le 0.001~\rm cm^{-3}$,
 $m_{\rm tot} \ge 3\times 10^6 M_{\odot}$, and $d_{\rm cl, max}\le2$ kpc.  These findings are in accord with a picture in which the cool inner CGM is dominated by numerous low-mass cloudlets ($m_{\rm cl}\lesssim 100 M_{\odot}$) with a filling factor $\lesssim1\%$.

This constraint is well below the mass scale that can currently be resolved in cosmological zoom simulations, even in those which pursue enhanced spatial refinement in circumgalactic regions \citep[e.g.,][]{Hummels19,Vandevoort19,Peeples19,Suresh19,Ramesh23}.
Given these limitations and the computational cost of further resolution enhancements,
we propose that \textsc{CloudFlex} be used in tandem with cosmological zoom simulations
as a sub-grid model for material in the cool phase.  This could in principle reduce
their resolution limits by orders of magnitude.  In addition, given that the inclusion of different physical effects in idealized hydrodynamic simulations (e.g., radiative cooling, self-shielding, thermal conduction, magnetic fields) can lead to significantly different predictions for cold cloud survival and morphology \citep[e.g.,][]{Armillotta2016,Li20,Tan21,gronke22,Abruzzo2023,Das23,Hidalgo-Pineda23}, a sub-grid implementation of \textsc{CloudFlex} will allow for a flexible exploration of the implications of these predictions on absorption-line observables.  Moreover, by further integrating the \textsc{trident} software into our modeling framework, it will be straightforward to extend our predictions to other commonly-observed absorption transitions.

These predictions may then be tested against datasets built from lensed QSOs or other extended background sources.  Such measurements probe the spatial and kinematic variation in absorption-line profiles across both very small scales (${\lesssim} 1$ kpc; \citealt{Rauch1999,Rauch2001,Rauch2002,Peroux2018,Rubin2018c,rubin18a,augustin21}) as well as across the scales of galaxy halos \citep{Chen2014,Zahedy2016,Lopez2018,Lopez2020,Tejos2021,Afruni2023}.  By confronting these datasets with our modeling framework, we may begin to constrain the mass distribution of the cool phase and the turbulent properties of both the cool material and the hot medium in which it is embedded. We likewise anticipate that \textsc{CloudFlex} will be an important tool for designing new datasets that improve these constraints.


\begin{acknowledgments}
This research was supported in part by the National Science Foundation under Grant No.\ NSF PHY-1748958.  CBH is supported by NSF grant AAG-1911233, and NASA grants 80NSSC23K1515, HST-AR-15800, HST-AR-16633, and HST-GO-16703.  KHRR acknowledges partial support from NSF grants AST-1715630 and AST-2009417.  EES acknowledges support from  NASA grants 80NSSC21K0271, 80NSSC22K0720, HST-AR-16633, and the David and Lucile Packard Foundation. The authors are most grateful for insightful input from Kirill Tchernyshyov, Peng Oh, Jessica Werk, Iryna Butsky, Freeke van de Voort, Sam Ponnada, Phil Hopkins, Zach Hafen, and Fakhry Zahedy.
\end{acknowledgments}

%



\software{\textsc{Trident} \citep{hummels17}, \textsc{unyt} \citep{Goldbaum2018}, \textsc{yt} \citep{turk11}}

\bibliography{clouds}{}
\bibliographystyle{aasjournal}



\appendix

\section{The Relation Between Cloudlet Complex Parameters $\alpha$, $\zeta$, Metallicity, $v_{\rm max}$, and $\beta$ and \ion{Mg}{2} Absorption}
\label{app:other_params}

\begin{figure*}[ht]
\centering
\includegraphics[width=\textwidth,trim={0cm 0.5cm 0cm 0.5cm},clip]{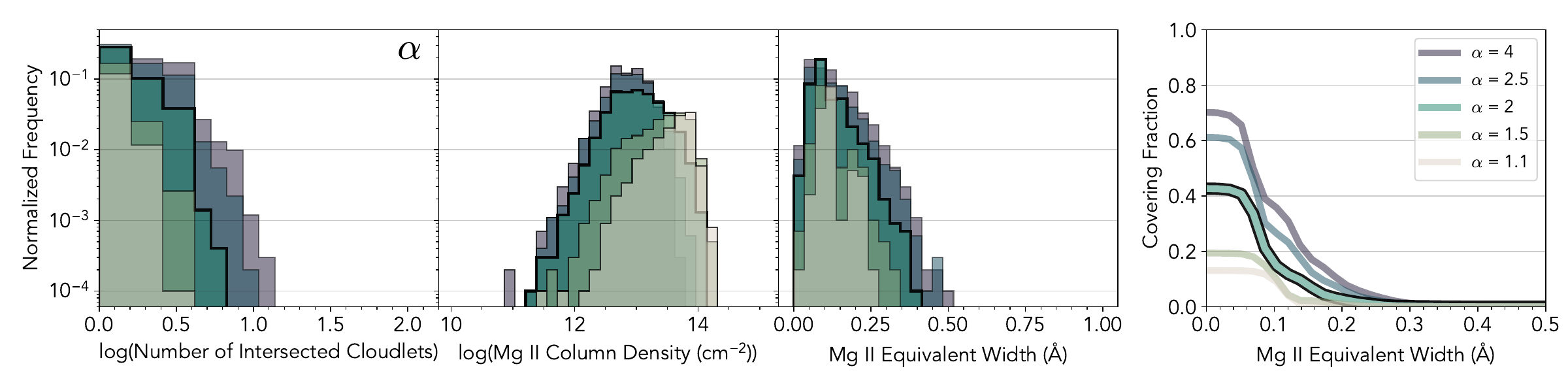}
\includegraphics[width=\textwidth,trim={0cm 0.5cm 0cm 0.5cm},clip]{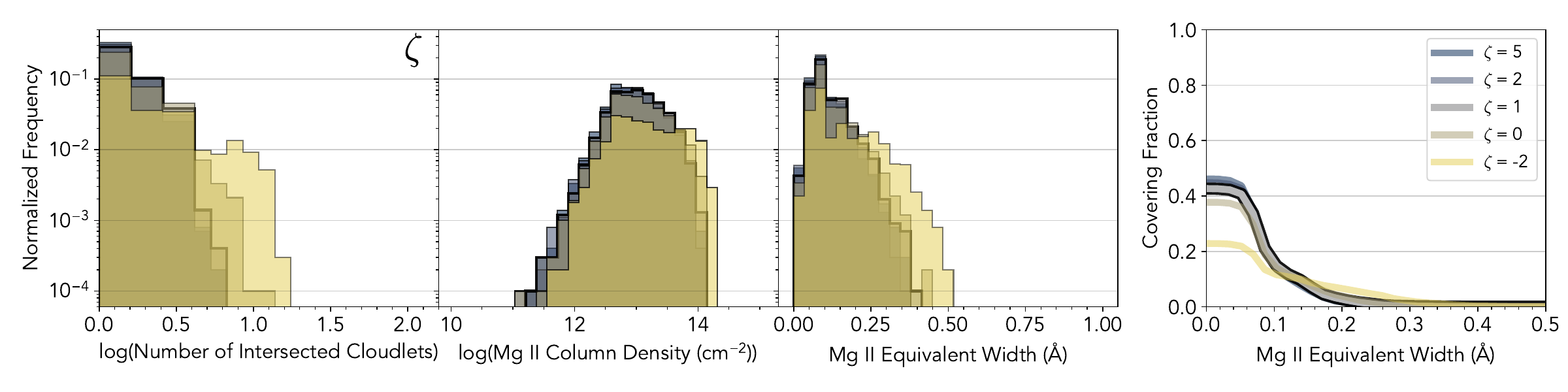}
\includegraphics[width=\textwidth,trim={0cm 0.5cm 0cm 0.5cm},clip]{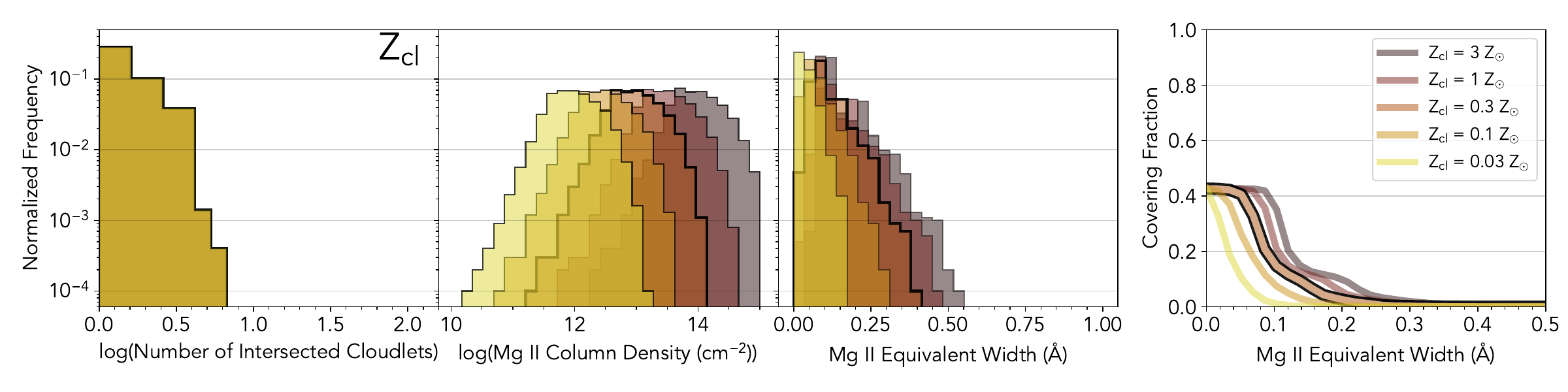}
\includegraphics[width=\textwidth,trim={0cm 0.5cm 0cm 0.5cm},clip]{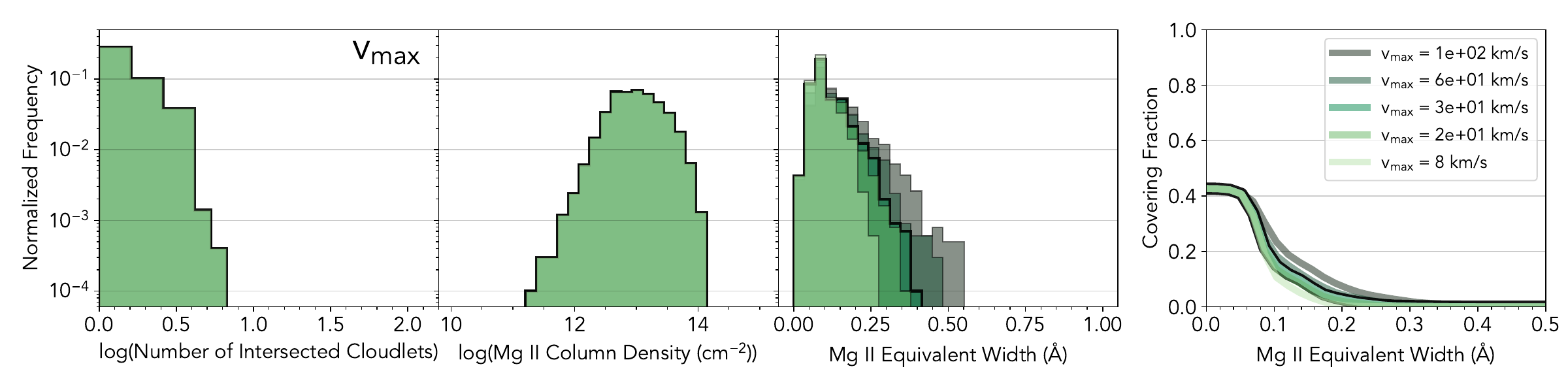}
\includegraphics[width=\textwidth,trim={0cm 0.5cm 0cm 0.5cm},clip]{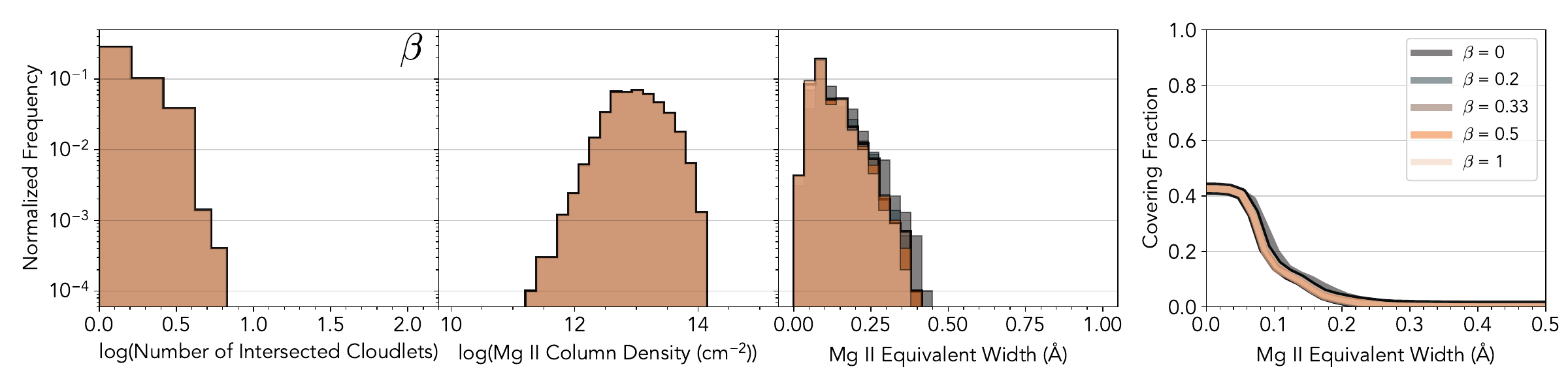}
\caption{Distributions of cloudlet complex \ion{Mg}{2}-absorption properties for models with varied input parameters not shown in Figure~\ref{fig:histograms}. Each row shows results for variations in a single model parameter ($\alpha$, $\zeta$, $Z_{\rm cl}$, $v_{\rm max}$, and $\beta$, from top to bottom), with all other parameters fixed at their fiducial values. From left to right, each column shows distributions of the number of cloudlets intercepted per sightline, \ion{Mg}{2} column density ($\NMgII$), and \ion{Mg}{2} equivalent width $W_{2796}$, with the fiducial model outlined in black.  Each distribution is normalized by the number of sight lines sampled, and we have excluded $W_{2796}$ values of 0 \AA\ from all $W_{2796}$ distributions for clarity.  The right-most panels show the covering fraction of \ion{Mg}{2} absorption for equivalent widths greater than the values indicated on the $x$-axis ($f_{\rm C}^{\rm cc}({>}W_{\rm X})$).} 
\label{fig:appendix}
\end{figure*}

Here we demonstrate how variations in the \textsc{CloudFlex} model parameters $\alpha$ (the power-law slope of the cloudlet mass distribution), $\zeta$ (the power-law slope of the cloudlet radial distance distribution), $Z_{\rm cl}$ (the cloudlet metallicity), $v_{\rm max}$ (the cloudlet turbulent velocity normalization), and $\beta$ (the slope of the turbulent power spectrum) affect observed \ion{Mg}{2} absorption properties. 
Figure~\ref{fig:appendix} shows distributions of \ion{Mg}{2} observables for cloudlet complexes 
generated with values of $\alpha$ over the range $1.1 \le \alpha \le 4$ (top); for values of $\zeta$ over the range $-2 \le \zeta \le 5$ (second row); for metallicities over the range $0.03Z_{\odot} \le Z_{\rm cl} \le 3Z_{\odot}$ (third row); for $v_{\rm max}$ values over the range $8~\mathrm{km~s^{-1}} \le v_{\rm max} \le 100~\mathrm{km~s^{-1}}$ (fourth row); and for values of $\beta$ over the range $0 \le \beta \le 1$ (bottom row).  As described in Section~\ref{sec:analysis}, all other parameters are set at their fiducial values.  

We find that variations in $\alpha$, the cloudlet mass distribution power law slope, have qualitatively similar effects on the observed $W_{2796}$ distributions as does changing the value of $m_{\rm cl,min}$.  A steeper slope ($\alpha > 2$) increases the relative numbers of low-mass cloudlets, thereby increasing the incidence of absorption over all equivalent widths $W_{\rm X}$.  Variations in $\zeta$, the cloudlet radial distance distribution power law slope, have only very minor effects on the observed $W_{2796}$ distributions for values $\zeta \ge 0$.  However, the value $\zeta=-2$ forces many cloudlets to be concentrated toward the inner part of the complex (as described in detail in Section~\ref{subsubsec:clobber}), such that a relatively large fraction of sight lines intercept $\sim 10-15$ cloudlets.  At the same time, the outer parts of the complex are sparsely populated, resulting in numerous ``empty'' sight lines.  

Variations in cloudlet metallicity affect the observed \ion{Mg}{2} column density and $W_{2796}$ distributions in a straightforward way, with increasing values of $Z_{\rm cl}$ yielding larger values of both observed quantities.  

Variations in the parameters regulating the turbulent properties of the cloudlets have no effect on the distributions of the numbers of intersected cloudlets and \ion{Mg}{2} column densities, as expected.  Changes in the value of $v_{\rm max}$, the factor used to normalize the cloudlet velocity distribution, increase the maximum observed $W_{2796}$ from $\approx 0.25$ \AA\ to $\approx 0.6$ \AA\ over the range we explore ($8-100~\rm km~s^{-1}$). However, the low-$W_{2796}$ part (at $W_{2796} < 0.15$ \AA) of the distribution is unaffected, as such weak absorbers must arise primarily from sight lines that intercept only a single cloudlet.  Variations in the turbulent power law slope, $\beta$, have a yet weaker effect on the observed $W_{2796}$ distribution, implying that measurements of this quantity from single sight line studies are unlikely to yield constraints on this parameter.  Multi-sightline datasets that enable study of the cross-correlation of absorption profiles \citep[e.g.,][]{Rauch2001} as a function of separation may be needed to improve our understanding of the turbulent properties of the hot CGM.  Based on the analysis presented in Figure~\ref{fig:spec_demo} and Section~\ref{subsec:analysis_mclmin}, we posit that recent work assessing the relation between the non-thermal broadening of low-ionization metal lines and the sizes of the absorbing clouds \citep{Chen23} is likely constraining a combination of internal and inter-cloudlet turbulence, 
due to blending of cloudlet absorption profiles below the resolution limit of the dataset (with FWHM $\approx 5-18~\rm km~s^{-1}$). 

In summary, we conclude that variation of the model parameters $\zeta$, $v_{\rm max}$, and $\beta$ produce changes in the overall distributions of \ion{Mg}{2} observables that are less significant than those discussed in Section~\ref{sec:analysis}.  Changes in the parameter $\alpha$ yield changes to our distributions of \ion{Mg}{2} observables that are very similar to those resulting from changes in $m_{\rm cl,min}$.


\end{document}